# Structure-Property Linkage in Shocked Multi-Material Flows Using A Level-Set Based Eulerian Image-To-Computation Framework

*S Roy, N Rai, O Sen and H.S. Udaykumar**

Department of Mechanical Engineering

The University of Iowa, Iowa City, IA-52242

---

**ABSTRACT**

Morphology and dynamics at the meso-scale play crucial roles in the overall macro- or system-scale flow of heterogeneous materials. In a multi-scale framework, closure models upscale unresolved sub-grid (meso-scale) physics and therefore encapsulate structure-property (S-P) linkages to predict performance at the macro-scale. This work establishes a route to structure-property linkage, proceeding all the way from imaged micro-structures to flow computations in one unified levelset-based framework. Levelsets are used to: 1) Define embedded geometries via image segmentation; 2) Simulate the interaction of sharp immersed boundaries with the flow field; and 3) Calculate morphological metrics to quantify structure. Meso-scale dynamics is computed to calculate sub-grid properties, i.e. closure models for momentum and energy equations. The structure-property linkage is demonstrated for two types of multi-material flows: interaction of shocks with a cloud of particles and reactive meso-mechanics of pressed energetic materials. We also present an approach to connect local morphological characteristics in a microstructure containing topologically complex features with the shock response of imaged samples of such materials. This paves the way for using geometric machine learning techniques to associate imaged morphologies with their properties.

------------------------------------------------------------------------------------------------------------------------

## 1 INTRODUCTION

In many engineering applications, the observed macro-scale response of a material is influenced, even controlled, by the physics at the scale of its microstructure (hereafter called the meso-scale). Meso-scale dynamics is especially important when heterogeneous materials (comprising mixed phases or materials) are subject to forces that result in large deformations culminating in instability, failure, or explosive energy release. Examples of such phenomena include hotspot-initiated detonation in heterogeneous energetic materials [2-4], evolution of particle-laden gases under blast conditions [5], failure of composites from delamination [6], cavitation-induced damage in liquid-vapor mixtures [7], phase transitions [8], failure at interfaces (debonding/delamination/fracture/crack propagation) [9, 10], defect growth [11] and damage evolution [12, 13]. Computational modelers of such phenomena must contend with the disparity of spatial and temporal scales —between the observable system (macro-) scale and the underlying microstructure (meso-scale).

Accurate modeling and prediction of the macro-scale dynamics of heterogeneous, micro-structured materials requires upscaling of the unresolvable (sub-grid) meso-scale physics. The upscaled information is incorporated in closure models, which encapsulate the physio-chemical response at the meso-scale; closure models are supplied as source or forcing terms in macro-scale governing equations. One can then use the closure models to predict the observed macro-scale response to shock loads (i.e. performance). Therefore, structure-property-performance [14, 15] linkages connect microstructural morphology (i.e. structure) to the observed response of the material. Here, we develop a levelset-based framework for developing S-P linkages using high fidelity meso-scale simulations of micro-structured materials. This paper builds on a body of previous work that connects the meso-scale to the macro-scale by developing surrogate models to capture sub-grid properties [16-18].

The process of establishing the structure-property-performance linkage is illustrated in Figure 1. The route shown in Figure 1 centers on meso-scale simulations, which resolve microstructural features and model meso-scale physics with high fidelity [16, 19]. These simulations employ computational mechanics solvers to calculate meso-scale quantities of interest (QoIs, $\mathbb{Q}$), which are then expressed as surrogate models [20] in a suitable parametric space. The $n_p$-dimensional set of parameters may consist of loading conditions, morphological metrics, material model parameters, chemical reaction rates and so on [17]. In engineering problems, the number of parameters $n_p$ can be very large; the surrogate model construction therefore is plagued by the curse of dimensionality. Circumventing curse of dimensionality is quite challenging, but strategies can be devised through variable selection methods to reduce the dimensions of the parameter space [21]; another solution can be the use of multi-fidelity methods for surrogate modeling [22].

The particular focus of this work is the simulation of shocked heterogeneous materials. In this context, the present multi-scale structure-property (S-P) linkage methodology has three interacting components, as illustrated in Figure 1:

1) A high-fidelity computational solver [19, 23] to calculate the meso-scale flow field, outlined in the Appendix A1.

2) Procedures to extract the meso-scale QoIs for surrogate modeling are outlined in Appendix A2 and A3 [16]

3) A morphometry capability that computes morphological metrics is described in Section 2.1.

This third aspect is the central contribution of this paper. The first two have been presented in detail in previous work and are only briefly touched upon here for the sake of completeness.

### 1.1 Representation and quantification of morphology

A popular approach to represent morphology in a wide range of fields is by using implicit fields, e.g. volume fractions [24], phase fields [25] or levelsets [26]; the embedded features are then picked out as low-dimensional iso-levels of these fields. In image-based modeling [27], levelsets are a natural choice. Images can be derived from a variety of modalities and the raw voxelated data is processed and segmented to obtain signed normal distance functions in Cartesian space[28]. Interfaces between materials in this representation are identified sharply as the zero-level contour on the voxel grid. In previous work, these intrinsic features of levelsets have been shown to provide an image-to-computation framework [28] to solve the high-speed dynamics of complex imaged meso-structures [29]. In other applications it was also used for simulation of video-to-computation applications in biological flows [28]. In an image-to-computation framework, levelsets offer several key advantages:

1. Levelsets rely on distance functions, which facilitates calculation of topological features such as curvature, interface area and the distance between embedded geometries.
2. The levelsets can conform to complex topology without the need to generate complicated meshes.
3. Dynamics, including fragmentation and coalescence, is naturally handled [10, 30].
4. Levelsets allow for sharp interface treatment of embedded objects [31, 32].
5. Extension to 3-dimensions is relatively straightforward since all operations are performed dimension-by-dimension on the levelset field [19].

Here we seek to utilize levelsets not only to represent and evolve interfaces but also to characterize the evolving morphology represented by the dynamic interfaces. In general, once the shapes of objects are defined by levelsets or other means, morphological metrics can be obtained in two broad categories to establish structure-property-performance linkages [33]: 1) Shape descriptors, and 2) Correlation functions. The former directly relate metrics to specific physical measures of the structure, for example, volume fractions, particle size and shape distributions, nearest neighbor distances etc. The latter provides statistical descriptions which may not be directly associated with physical characteristics, for example, n-point correlations [34] — $n = 0$ provides volume fraction, $n = 1$ provides the probability of finding a phase at a given location, and $n = 2$ provides descriptions of relative arrangement of features or clustering as a function of scale. Correlations at n>3 level are seldom used as they can be quite expensive to compute and may not be significant for structure-property correlations. In this work, we focus on developing shape descriptors to characterize morphology, as these are most directly relevant to the physics problems of interest herein.

Under the category of shape descriptors, the fields of morphometry [35, 36] and stereology [37] have produced a host of methods for morphology characterization, with mature codes dedicated to quantification of morphologies. The characterization of meso-structure of materials involves the quantitative description of the size, shape, numbers, and position of features (particles, void spaces, interfaces) within the specimen studied. Previous work in microstructural characterization comes from various applications, for example in characterizing bone structure [35, 36], soil pore networks [38], porous rocks [36], neuro-anatomy [37] and vasculature [39]. The useful microstructural quantities for each of these systems vary and correspondingly the morphometry techniques used also vary widely. For example, in trabecular bones, the load bearing capacity depends on the loading direction if the structure is anisotropic [40]. To quantify interfacial anisotropy, the mean intercept length metric for the bone and marrow spaces as a function of orientation is calculated, to characterize the dominant orientation in cancellous bone architectures [41]. An alternative method for calculating the interfacial anisotropy is the volume anisotropy proposed by [42], which is useful to obtain information on the overall orientation of the particles or voids in a microstructure [43]. In order to characterize microstructure of soil pore networks or retinal blood vessels where there is no constant orientation, measures such as tortuosity and connectivity are used [38, 39]. These tortuosity and connectivity measures are also correlated with gas diffusivity and air permeability for soil pore networks [38]. The methods and metrics of choice therefore depend on the micro-structure of the material and the desired QoI associated with the property to be quantified.

In this work, we develop methods to use levelsets to calculate a variety of shape descriptors of micro-structures. The use of levelset fields in characterizing morphology in this way has only received limited attention. [44] used levelsets for structural topology optimization, and for designing heterogeneous structures [45], while [46] obtained Minkowski functionals for surface structure characterization using levelsets. [47] used levelsets for polygonization of interface boundaries for mesh generation and to calculate the triple phase boundary curve and interface area. In the X-FEM context, [48] used the levelset representation of surfaces to simplify mesh generation for complex material interfaces and geometrical pre-processing. However, levelsets have not yet been used to comprehensively characterize the microstructure of heterogeneous materials, particularly for structure-property linkages. In this paper levelsets are used both for quantification of microstructural metrics and for performing mesoscale flow computations using a sharp interface treatment of the geometries [31, 32]. Using levelsets for both these applications facilitates the establishment of structure-property linkage in multi-material systems.

### 1.2 Organization of the paper

The levelset-based morphometry techniques are developed in this work by focusing on two computational mechanics problems [10, 23, 29-31, 49], viz.:

1. The evolution of particle-laden shocked flows, with application to blast and dusty gas flows [50]. Figure 1 indicates the way in which the structure-property linkage is established for this physics problem. Here the meso-structure involves clouds of micron-sized solid particles that are carried by the compressible gas medium. The structure to be quantified is fairly simple, consisting of clusters of cylindrical particles. The size, distribution and relative arrangement of solid particles at the meso-scale influence the closure laws (inter-phase mass, momentum and energy transfers) that describe the evolution of the overall cloud.
2. The reactive dynamics of imaged meso-structures of a commonly used pressed energetic material (HMX) [29]. Figure 1 indicates the approach to link property and structure in the case of the energetic material problem. The HMX crystals are pressed to high theoretical mean density (TMD) and void spaces exist between the crystals. The structure can vary from elongated void shapes to complex, sinuous and branched voids and cracks. The closure law in this case must quantify the sensitivity of the meso-scale sample as a function of the shape and distributions of voids in the pressed explosive.

The above two problems present and demand rather different types of morphology characterizations and therefore provide a testbed for developing the present levelset-based facility for structure-property linkage.

This remainder of the paper is organized as follows. The methods used to extract the structural metrics for both applications are shown in Section 2.1. Section 3.1 shows the verification of the methods. Finally, in Section 3.3 and 3.4, the S-P linkage for shocked gas particle flows and pressed HMX respectively are obtained. The paper ends with concluding remarks in Section 4. Appendix A1 outlines the methods used for solving the governing equations in the mesoscale simulations. Techniques used to obtain the QoIs for the particle-laden shocked flows and reactive dynamics of pressed HMX are described in appendix A2 and A3 respectively.

## 2 METHODS

### 2.1 Quantifying structure: Morphology characterization using levelset fields

Features or objects can be extracted from images by segmenting the greyscale intensity field acquired from X-ray, computed tomography (CT), scanning electron microscope (SEM) or ultrasound images, as described in previous work on image-to-computation techniques [28]. The active contour evolution approach [51] is used to extract geometries from images with varying levels of noise and intensity [27]. This method produces contours that are smoothed prior to segmentation using speckle-reducing anisotropic diffusion [52]. Figure 1 (a) shows the SEM image of a sample of HMX segmented through this procedure [53]. Figure 1 (b) shows the segmented image, where the levelset field values $\varphi(\boldsymbol{x}, t) = 0$ define the void interface in the sample.

Following segmentation, each object is numbered using an object tagging algorithm [54]. The microstructural features for the labeled objects can then be quantified and distribution functions can be generated for each metric of importance in the microstructure, as illustrated in Figure 1 (c). Within the computational domain shown in Figure 1 (d), the RVE is denoted by $\Omega$. Data to calculate meso-scale QoIs is collected in this RVE, to mitigate edge effects at domain boundaries. i, j, and k denote the grid indices in the Cartesian space and $\boldsymbol{x}$ is the location of grid points. A variety of shape descriptors are calculated from the levelset field, as described in the following sections.

### *2.1.1 Quantifying particle size, interface area and surface curvature*

The levelset field of an object with index l is denoted by the field $\varphi_l(\boldsymbol{x},t)$, the signed normal distance to interfaces defined on a Cartesian grid with nodes at locations $\boldsymbol{x}$. $\varphi_l(\boldsymbol{x},t) < 0$ in the interior and $\varphi_l(\boldsymbol{x},t) > 0$ outside the object. Volumes (area in 2D) of objects can be obtained from levelset fields by defining a Heaviside function:

$$\mathcal{H}_l(\boldsymbol{x},t) = \begin{cases} 0 & if\ \varphi_l(\boldsymbol{x},t) \leq 0 \\ 1 & if\ \varphi_l(\boldsymbol{x},t) > 0 \end{cases} \tag{1}$$

The volume (area in 2D) of an object is calculated from:

$$\Omega_l = \int_{\Omega} \big(1 - \mathcal{H}_l(\boldsymbol{x},t)\big) d\,\Omega \tag{2}$$

The interface area (perimeter in 2D) of an object is obtained by locating the zero levelset contour with sub-cell resolution [44]. Then, the interface area (perimeter in 2D) is calculated from:

$$\Gamma_l = \int_{\Omega} \delta_l(\boldsymbol{x},t) |\nabla \varphi_l(\boldsymbol{x},t)| d\,\Omega \tag{3}$$

Here $\delta_l(\boldsymbol{x},t)$ is a mollified Dirac delta function:

$$\delta_l(\boldsymbol{x},t) = \begin{cases} 0 & if\ \big||\varphi_l(\boldsymbol{x},t)|\big| > \varepsilon \\ \dfrac{1}{2\varepsilon}\left[1 + \cos\left(\dfrac{\pi\, ||\varphi_l(\boldsymbol{x},t)||}{\varepsilon}\right)\right] & if\ \big||\varphi_l(\boldsymbol{x},t)|\big| \leq \varepsilon \end{cases} \tag{4}$$

where $\varepsilon$ is the spread of the delta function.

The ratio of interface area (perimeter in 2D) to volume (area in 2D) denoted by $\ell_l$ is then:

$$\ell_l = \frac{\Gamma_l}{\Omega_l} \tag{5}$$

The normal vector is given by $\vec{n}(\boldsymbol{x},t) = \frac{\nabla \varphi_l(\boldsymbol{x},t)}{||\nabla \varphi_l(\boldsymbol{x},t)||}$. The curvature of the interface is calculated using the divergence of the unit normal vector:

$$\kappa_l(\boldsymbol{x},t) = \nabla \cdot \frac{\nabla \varphi_l(\boldsymbol{x},t)}{||\nabla \varphi_l(\boldsymbol{x},t)||} \tag{6}$$

The mean surface curvature for an object indexed l is calculated by integrating the curvature $\kappa_l(\vec{x}, t)$ along the interface $\Gamma_l$ of the object:

$$K_l = \frac{1}{\Gamma_l} \int_{\Gamma_l} |\kappa(\boldsymbol{x}, t)| \, d\Gamma_l \tag{7}$$

### 2.1.2 *Morphological metrics associated with object shapes*

To characterize the shapes of objects, the aspect ratio and orientation of the $l^{\text{th}}$ object can be obtained using the volume orientation method [43]. The volume orientation method is first used to obtain the local orientation and aspect ratios at each point in the Cartesian grid. To do this, line intercepts are sampled, radiating from all grid points inside the objects. The sampled intercepts are extended in orientations $\theta$ and $\theta + \pi$ until they cross the $\varphi_l(\boldsymbol{x}, t) = 0$ contour. Figure 2 (a) shows an illustration of the sampled intercepts $I\,(\boldsymbol{x}, \theta, t)$. The set of the sampled intercepts at a location $\boldsymbol{x}$ and oriented at various $\theta$ is defined as:

$$\boldsymbol{I}(\boldsymbol{x}, t) = \{I(\boldsymbol{x}, \theta, t) \mid 0 \leq \theta \leq \pi\} \tag{8}$$

The intercept with the maximum length in the set $\boldsymbol{I}(\boldsymbol{x}, t)$ is obtained as:

$$I^{\max}(\boldsymbol{x}, t) = \max(\boldsymbol{I}(\boldsymbol{x}, t)) \tag{9}$$

$I^{\max}(\boldsymbol{x}, t)$ is illustrated for a void belonging to Sample A in Figure 2(b). The local orientation $\omega(\boldsymbol{x}, t)$ at each grid point in the interior of the object is then defined as:

$$\omega(\boldsymbol{x}, t) = \{\theta | I\,(\boldsymbol{x}, \theta, t) = I^{\max}(\boldsymbol{x}, t)\} \tag{10}$$

Figure 2 (b) illustrates the way in which the local orientation $\omega(\boldsymbol{x}, t)$ is calculated at a point inside an object. The mean orientation for the $l^{\text{th}}$ object is calculated as:

$$\Theta_l = \frac{1}{\Omega_l} \int_{\Omega_l} \omega(\boldsymbol{x}, t) \, d\,\Omega_l \tag{11}$$

The local aspect ratio $\alpha(\boldsymbol{x}, t)$ is calculated based on the ratio of the length of the intercept $I^{\max}$ to the intercept $I^{\perp}$ which is perpendicular to $I^{\max}$ (Figure 2 (b)):

$$\alpha(\boldsymbol{x}, t) = \frac{I^{\max}(\boldsymbol{x}, t)}{I^{\perp}(\boldsymbol{x}, t)} \tag{12}$$

The mean aspect ratio for the $l^{\text{th}}$ object is calculated by taking the volume average of local aspect ratios:

$$\mathcal{A}_l = \frac{1}{\Omega_l} \int_{\Omega_l} \alpha(\boldsymbol{x}, t) \, d\,\Omega_l \tag{13}$$

### 2.1.3 *Characterizing microstructural anisotropy*

The presence of strong directionality or order in the arrangement of features in the microstructure can influence the dynamics of a heterogeneous material. This directionality in the material, i.e. microstructural anisotropy [55-57] has been characterized using the fabric tensor [40, 55, 58]. Using quantities such as mean intercept length [35, 58] and star length [59], the fabric tensor can be calculated from the levelset field, as described below.

*2.1.3.1 Star Length (SL) Method*

The star length method is a global anisotropy metric used to quantify the overall directionality of features or objects in the microstructure [42, 43, 58, 60]. Figure 2 (a) illustrates this method for characterizing anisotropy due to voids in a heterogeneous energetic material sample. The white dot indicates a point inside the object at a spatial location $\boldsymbol{x}$ from which lines are sampled in orientations $\theta$. The yellow dots indicate the intersection points between the lines originating from $\boldsymbol{x}$ and the object boundary ($\varphi_{\mathrm{l}}(\boldsymbol{x}, t) = 0$ contour). The intersection points are recorded by tracing along lines from $\boldsymbol{x}$ until the levelset values corresponding to the points in the line flip signs. The length of the line intercept extended from $\boldsymbol{x}$ in the directions $\theta$ and $\theta + 180°$ proscribed by its intersection with the object interface is denoted by $I(\boldsymbol{x}, \theta)$.

Line intercepts $I(\boldsymbol{x}, \theta)$ are traced at all spatial locations $\boldsymbol{x}$ within the objects. At each location $\boldsymbol{x}$, intercepts are traced in $N$ directions with angular intervals of $\Delta\theta$. The star length $L_{\mathrm{S}}(\theta)$ is calculated using the following equation:

$$L_{\mathrm{S}}(\theta) = \frac{1}{\Omega} \int_{\Omega} I(\boldsymbol{x}, \theta)\, d\,\Omega \tag{14}$$

The $L_{\mathrm{S}}(\theta)$ values are plotted as a polar plot and an ellipse is fit to the $L_{\mathrm{S}}(\theta)$ using the method described below. The ellipse fitting method can be used to obtain the anisotropy tensor $\mathcal{M}$, anisotropy index $A_{\mathrm{index}}$, and anisotropic orientation $\Psi$ [35, 61, 62].

The ellipse is fit using a least-squares approximation to the contour of $L_{\mathrm{S}}(\theta)$ data. Figure 2 (c) shows the illustration of the ellipse fitting process in the x-y plane. The grey dots indicate the points indexed k that are obtained from the $L_{\mathrm{S}}(\theta)$, where the x and y coordinates of each of the points are given by $x_{\mathrm{k}} = L_{\mathrm{S}}(\theta_{\mathrm{k}}) cos\theta_{\mathrm{k}}$, and $y_{\mathrm{k}} = L_{\mathrm{S}}(\theta_{\mathrm{k}}) sin\theta_{\mathrm{k}}$, $\mathrm{k} = 1, 2, \dots, N$. The general equation of the fitting ellipse in x-y coordinates is:

$$A\, x_{\mathrm{k}}^{2} + B\, y_{\mathrm{k}}^{2} + C\, x_{\mathrm{k}} y_{\mathrm{k}} + D = 0 \tag{15}$$

This is expressed as:

$$\mathbf{A}\, \boldsymbol{\xi}_{\mathrm{k}} = 0 \tag{16}$$

where the coefficient vector $\boldsymbol{A} = [A\ B\ C\ D]^{\mathsf{T}}$, and $\boldsymbol{\xi}_{\mathrm{k}} = [\, x_{\mathrm{k}}^{2}\ y_{\mathrm{k}}^{2}\ x_{\mathrm{k}} y_{\mathrm{k}}\ 1]$. $\boldsymbol{A}$ is then determined using a least-squares fit to the sequence of points $(x_1, y_1)$, … , $(x_{\mathrm{N}}, y_{\mathrm{N}})$. An eigenvalue problem is constructed using the matrix $\mathcal{M}$ given by:

$$\mathcal{M} = \frac{1}{\mathrm{N}} \sum_{\mathrm{k}=1}^{\mathrm{N}} \boldsymbol{\xi}_{\mathrm{k}} \boldsymbol{\xi}_{\mathrm{k}}^{\mathsf{T}} \tag{17}$$

Then, the following eigenvalue problem is solved to obtain the coefficient vector $\boldsymbol{A}$ [63]:

$$\mathcal{M}\boldsymbol{A} = \lambda \boldsymbol{A} \qquad \text{subject to } \|\boldsymbol{A}\| = 1 \tag{18}$$

The smallest eigenvalue $\lambda$ is calculated by minimizing $J$, i.e. the sum of squared distances [64]:

$$J = \frac{1}{\mathrm{N}}\sum_{\mathrm{k}=1}^{\mathrm{N}}(\boldsymbol{\xi}_{\mathrm{k}}, \boldsymbol{A})^2 = \frac{1}{\mathrm{N}}\sum_{\mathrm{k}=1}^{\mathrm{N}} \boldsymbol{A}^{\mathrm{T}}\boldsymbol{\xi}_{\mathrm{k}}\boldsymbol{\xi}_{\mathrm{k}}{}^{\mathrm{T}}\boldsymbol{A} = \boldsymbol{A}^{\mathrm{T}}\mathcal{M}\boldsymbol{A} \tag{19}$$

The coefficient vector $\boldsymbol{A}$, can be used to calculate the anisotropy index $A_{\mathrm{index}}$ as the ratio:

$$A_{\mathrm{index}} = {}^{B}/_{A} \tag{20}$$

The orientation of the ellipse $\Psi$ can be obtained from:

$$\Psi = \frac{1}{2}tan^{-1}\left(\frac{C}{A-B}\right) \tag{21}$$

### *2.1.4 Quantifying spatial patterns using levelsets*

The response of a heterogeneous material containing particles or voids in an otherwise uniform medium will depend on the distribution of the embedded objects. The objects may be spatially distributed in regular/ random or uniform/clustered fashion. There are various statistical methods to characterize the spatial distribution of objects [65, 66]. In this work, levelsets are used to calculate measures that characterize the spatial distribution of objects using inter-particle, nearest-neighbor and nearest-interface distances. These measures provide information on clustering or aggregation, regularity of spatial distribution of objects or complete spatial randomness (CSR) [67] and relative proximity of neighboring objects.

#### *2.1.4.1 Calculating nearest neighbor distance*

The nearest neighbor distribution function measures the proximity of object centroids with respect to each other. The centroid of an object numbered l is obtained from its levelset field $\varphi_{\mathrm{l}}(\boldsymbol{x})$. The interior of each object (where $\varphi_{\mathrm{l}}(\boldsymbol{x})$ <0) is decomposed into Cartesian grid cell cubical volumes at each spatial location $\boldsymbol{x}$ in the grid. The centroids of the individual objects $\boldsymbol{c}_{\mathbf{l}}$, is then calculated in a straightforward way:

$$\boldsymbol{c}_{\mathbf{l}} = \frac{1}{\Omega_{\mathrm{l}}}\int_{\Omega_{\mathrm{l}}} \boldsymbol{x}\, d\Omega_{l} \tag{22}$$

The nearest neighbor distance is obtained by first calculating the inter-centroid distance $d_{\mathrm{lk}}$. $d_{\mathrm{lk}}$ is the Euclidean distance between centroids $\boldsymbol{c}_{\mathbf{k}}$ and $\boldsymbol{c}_{\mathbf{l}}$ of objects k and l respectively. The nearest neighbor distance from the $\mathrm{l}^{\mathrm{th}}$ object $\mathbb{D}_{\mathrm{l}}$ is then:

$$\mathbb{D}_{\mathrm{l}} = min\{d_{\mathrm{lk}} | 1 \leq \mathrm{k} \leq N_{\mathrm{obj}}, \mathrm{l} \neq \mathrm{k}\} \tag{23}$$

where $n_{\mathrm{obj}}$ is the total number of objects in the domain.

### 2.1.4.2 *Calculating nearest interface distance*

The nearest interface distance measures the proximity of object interfaces from each other and is an important metric to evaluate collisions between objects [68] or to apply tractions due to friction [69]. Since levelsets provide signed normal distances from particle interfaces, they can be used directly to calculate inter-interface distances between objects. For a system with object index l, separate levelset fields $\varphi_{\mathrm{l}}$ are generated for each object.

Figure 3 shows a schematic with the levelset field $\varphi_{\mathrm{l}}$ for an object (in this case a void) within an energetic material sample. Superposed on $\varphi_{\mathrm{l}}$ are the interfaces (0-levels) of all the other k objects (shown in black) where $\mathrm{k} = 1, \ldots, n_{\mathrm{obj}}$ and $\mathrm{k} \neq \mathrm{l}$. The shortest possible interface distance between object l and k is calculated by taking the minimum of the $\varphi_{\mathrm{l}}$ at the boundary of the k objects, i.e.:

$$\mathbb{i}_{\mathrm{lk}} = min\{\varphi_{\mathrm{l}}(\boldsymbol{x}, t) \ \forall \ \boldsymbol{x} \mid \varphi_k(\boldsymbol{x}, t) = 0, \mathrm{l} \neq \mathrm{k}\} \tag{24}$$

Figure 3 shows $\mathbb{i}_{\mathrm{l1}}$, $\mathbb{i}_{\mathrm{l2}}$, $\mathbb{i}_{\mathrm{l3}}$, and $\mathbb{i}_{\mathrm{l4}}$ indicating the inter-interface distance between object l and objects with index $\mathrm{k} = 1, \ldots, 4$. The yellow dots in the figure show the points on the interfaces of objects numbered 1 through 4 at which the value of the $\varphi_{\mathrm{l}}$ is minimum.

The nearest interface distance $\mathbb{I}_{\mathrm{l}}$ for object l is obtained by taking the minimum of the interface distances from object l to all the other objects, i.e.,

$$\mathbb{I}_{\mathrm{l}} = min\left\{\mathbb{i}_{\mathrm{lk}} \mid 1 \leq \mathrm{k} \leq n_{\mathrm{obj}}, \mathrm{l} \neq \mathrm{k}\right\} \tag{25}$$

The $\mathbb{I}_{\mathrm{l}}$ for all the $n_{\mathrm{obj}}$ number of objects can be plotted as the nearest interface distribution.

### 2.1.4.3 *Calculating tortuosity using levelsets*

By definition, tortuosity is the ratio of the path length of the curve between two points to the length of the straight line segment connecting the two points [70]. For any system with an arrangement of particles, tortuosity can be measured by identifying the medial or topological skeleton and their pathways. In the current framework, the levelset field of the system of particles $\varphi(\boldsymbol{x}, t)$ is used to extract the medial skeleton. For example, the levelset field $\varphi(\boldsymbol{x}, t)$ for a particle cluster is shown in Figure 4 (a); the levelset field outside the particles rises to peak values. The locus of the peak values of $\varphi(\boldsymbol{x}, t)$ can be detected by taking the Laplacian $\boldsymbol{\nabla}^2\varphi(\boldsymbol{x}, t)$ of the levelset field in the regions outside the particle. The Laplacian of the levelset field $\boldsymbol{\nabla}^2\varphi(\boldsymbol{x}, t)$ has peak values on the medial skeleton. Thresholding of the $\boldsymbol{\nabla}^2\varphi(\boldsymbol{x}, t)$ field gives the medial skeleton. For the case shown, the threshold value used is 0.04. The medial skeleton obtained through this method can be multiple pixels thick. To reduce the skeleton to the width of a single pixel, a thinning algorithm is used [71].

The skeleton obtained from the levelset field is used to obtain the tortuosity for a cluster of particles. Figure 4 (b) shows the skeleton for the particle clusters. The shortest distance between points $A_{\mathrm{xi}}$ and $B_{\mathrm{xi}}$ traversed through the skeletal path is denoted by $d_{\mathrm{xi}}$, where subscript x denotes that the path is in x direction and i denotes the path number. The line segment connecting points $A_{\mathrm{xi}}$ and $B_{\mathrm{xi}}$ is denoted as $d_{\mathrm{xi}}^{\mathrm{min}}$ and illustrated in Figure 4 (d).Therefore, tortuosity for path number i in the x direction is given by the following equation.

$$T_{\mathrm{xi}} = \frac{d_{\mathrm{xi}}}{d_{\mathrm{xi}}^{\mathrm{min}}} \tag{26}$$

The tortuosity values $T_{\mathrm{xi}}$ for all the possible paths in the x direction where $\mathrm{i} = 1, \dots, m_{\mathrm{x}}$ are obtained. $m_{\mathrm{x}}$ is the total number of possible paths in x direction. Therefore, the average tortuosity in $T_{\mathrm{x}}$ in x direction is given by:

$$T_{\mathrm{x}} = \frac{1}{m_{\mathrm{x}}} \sum_{\mathrm{i}=1}^{m_{\mathrm{x}}} T_{\mathrm{xi}} \tag{27}$$

Similarly, the average tortuosity in the y-direction $T_{\mathrm{y}}$ is calculated as:

$$T_{\mathrm{y}} = \frac{1}{m_{\mathrm{y}}} \sum_{\mathrm{i}=1}^{m_{\mathrm{y}}} T_{\mathrm{yi}} \tag{28}$$

where $m_{\mathrm{y}}$ is the total number of paths in the y direction and $T_{\mathrm{yi}}$ the tortuosity of the $\mathrm{i}^{\mathrm{th}}$ path.

The notion of tortuosity is demonstrated by considering two types of arrangements of cylinders, i.e. inline and staggered. Figures 4 (c) and (c) shows the medial axis or topological skeletons of the cylinders [72]. From visual inspection it can be noted that the length of medial skeleton path $A - B$ is greater for the staggered arrangement (Figure 4 (c)) compared to the inline arrangement (Figure 4 (c)). Therefore the path $A - B$ for staggered arrangement has a higher tortuosity. Fluid particles negotiate longer paths while flowing over the staggered configuration, which has implications for vorticity generation, drag and scalar mixing in the particle clusters.

#### *2.1.4.4 Tessellation and Voronoi cells using levelsets*

Voronoi tessellations are often used in analyzing spatial point patterns[67, 73]. In Voronoi tessellation, the overall domain is sub-divided into smaller cells such that each cell contains a particle. Each Voronoi cell boundary is equidistant from the nearest particles on either side of the boundary; the generation of Voronoi cells is facilitated by the levelset field, which contains the normal distance information to the closest object interface. First, the Laplacian of the levelset field $\boldsymbol{\nabla}^2 \varphi(\boldsymbol{x}, t)$ is used to obtain the medial skeleton for a cluster of particles as shown in Figure 4 (b). The medial skeleton divides the domain directly into Voronoi cells as shown in Figure 7 (a) and (b). The regions inside the Voronoi cell edges are numbered separately such that each particle in the domain is contained within a Voronoi cell and the total number of Voronoi cells is equal to the total number of objects $n_{\mathrm{obj}}$ (see Figure 7 (a) and (b)).

The size (area in 2D and volume in 3D) of the Voronoi cells $\Omega_{\mathrm{l}}^{\mathrm{vor}}$ is calculated by summing the total number of pixels within each cell. The subscript l denotes the Voronoi cell corresponding to the particle numbered l. The distribution function of the Voronoi cell size $\Omega_{\mathrm{l}}^{\mathrm{vor}}$ is used to study regularity, clustering, and homogeneity in a spatially distributed particle system [74].

## 3 RESULTS

The methods for structure quantification described above and property evaluation from meso-scale simulation techniques described in Appendix A1-3 are applied to two systems in this section, viz. particle-laden gas flows and pressed porous solid HE materials. First, the levelset-based structural metric calculations are verified for canonical geometries. This is followed by the quantification of QoIs to be extracted from the meso-scale calculations for the two systems. Finally the overall methodology is applied to establish structure-property linkages for the two systems studied.

### 3.1 Verification of structural metrics

Interface area and curvature: The calculation of object interface area $\Gamma_{\mathrm{l}}$ from the levelset field, as explained in Section 2.1.1, is applied to a cylinder of diameter $d_{\mathrm{cyl}} = 0.5\ units$. The delta function is obtained based on Equation (4) for the cylinder and used for $\Gamma_{\mathrm{l}}$ calculation, the spread of the numerical delta function $\varepsilon = 12 * \Delta x$, where the grid spacing $\Delta x = 0.004$. The $\Gamma_{\mathrm{l}}$ calculated for this circle using Equation (3) is $1.5700\ units$, compared to the analytical value ($1.5708\ units$). The mean curvature for the circle is obtained based on Equation (7) $K_{\mathrm{l}} = 4.036\ units^{-1}$, which corresponds to an error of 0.009% compared to the analytical value of mean surface curvature ($4.0\ units^{-1}$). Thus the procedure in section 2.1.1 computes the interface perimeter $\Gamma_{\mathrm{l}}$ and curvature $K_{\mathrm{l}}$ accurately.

Star length (SL) and anisotropy: Verification of the star length ($SL$) method (Section 2.1.3.1) is done for a circle of diameter $d_{\mathrm{cyl}} = 0.5$ as shown in Figure 5 (a). The polar plot of the star length $L_{\mathrm{S}}(\theta)$, obtained using Equation (14), is shown with purple markers in Figure 5 (c). The analytically calculated expression of the star length $L_{\mathrm{S}}(\theta)$ for circles is $L_{\mathrm{S,exact}} = \frac{8\, d_{\mathrm{cyl}}}{3\,\pi}$ for all $\theta$ [59]. For $d_{\mathrm{cyl}} = 0.5$, the analytical value of star length is calculated as $L_{\mathrm{S,exact}} = 0.4244$; the average calculated $L_{\mathrm{S}}(\theta) = 0.4251$ (i.e. using Equation (14)), with a low standard deviation of 0.039%. The $L_{\mathrm{S}}(\theta)$ polar plot is also calculated for the ellipse shown in Figure 5 (d). The ellipse fit to the $L_{\mathrm{S}}(\theta)$ values using Equations (15) - (21) is shown in Figure 5 (f); from the ellipse fit $A_{\mathrm{index}} = 1.9969$ which is close to the ratio $d_{\mathrm{maj}}/d_{\mathrm{min}} = 2$ for the ellipse in Figure 5 (d). Hence, the anisotropy characterization method using the star length gives accurate results for the case of a circle and an ellipse

Aspect ratio and orientation of objects: A microstructure generation code [75] was used to generate a system of spatially distributed ellipses shown in Figure 6 (a), for verification of the object aspect ratio $\mathcal{A}_{\mathrm{l}}$ and orientation $\Theta_{\mathrm{l}}$ calculated using methods described in Section 2.1.2. The field of ellipses was generated such that the mean orientation was 90° with a standard deviation of 10°; the ratio of major axis to the minor axis of the ellipse were kept constant at $d_{\mathrm{maj}}/d_{\mathrm{min}} = 2$. The $\mathcal{A}_{\mathrm{l}}$ calculated for individual ellipses in the generated field of ellipses is shown in Figure 6 (b). The $\mathcal{A}_{\mathrm{l}}$ for all the ellipses have values ranging from 1.81 to 2.22, except for ellipses falling on the domain edges. Figure 6 (c) shows the plot for the $\Theta_{\mathrm{l}}$ distribution for all the ellipses in the field. The mean $\Theta_{\mathrm{l}}$ value calculated from the distribution plot is $90_{o}$ with a standard deviation of 10°. This is in good agreement with the mean and standard deviation of the orientation and aspect ratio values used as input for generating the field of ellipses [75].

Voronoi cell size ($V_{\mathrm{l}}^{\mathrm{vor}}$) distribution: The effect of spatial arrangement of a field of particles on Voronoi cell size distribution is shown in Figure 7 for two different arrangements. These arrangements are generated using particles of diameter $d_{\mathrm{cyl}} = 0.02$ and particle volume fraction of 10% within a domain of size $1 \times 1$. In the first case, the particles were placed randomly in the domain with the constraint that the inter-centroid distance $d_{\mathrm{kl}} > 0.025$ between any two particles k and l (Figure 7 (a)) [76]. In the second case, the constraint is $d_{\mathrm{kl}} > 0.045$ and the particles are placed randomly (Figure 7 (b)) [76].

The Voronoi cells shown in Figures 7 (a) and (b) were constructed for both cases using the method described in section 2.1.4. The first case, in Figure 7 (a), shows Voronoi cells of varying $\Omega_{\mathrm{l}}^{\mathrm{vor}}$ , compared to Figure 7 (b) where the $\Omega_{\mathrm{l}}^{\mathrm{vor}}$ have similar values. The standard deviation of the $\Omega_{\mathrm{l}}^{\mathrm{vor}}$ is shown in the $\Omega_{\mathrm{l}}^{\mathrm{vor}}$ distribution plots in Figures 7 (c) and (d). For the first case, the standard deviation in $\Omega_{\mathrm{l}}^{\mathrm{vor}}$ is much higher i.e. $0.34$ compared to the second case which shows a standard deviation of $0.19$. However, in both the cases the mean $\Omega_{\mathrm{l}}^{\mathrm{vor}} = 0.0029$. It can be concluded that there is a direct correlation between clustering and the standard deviation of the $\Omega_{\mathrm{l}}^{\mathrm{vor}}$ as suggested previously [77]. Therefore, the levelset based Voronoi cell size $\Omega_{\mathrm{l}}^{\mathrm{vor}}$ can be used to quantify inhomogeneities in particle distributions in many-particle systems.

### 3.2 Evaluating properties (meso-scale QoIs): the computational mechanics solver

The governing equations for compressible flow are solved for both the particle-laden gas and pressed energetic material problems. The governing equations are given in the Appendix followed by a brief description of the flow solver. Further details on the flow solver, including benchmarking and validation for a variety of physical problems can be obtained from previous works [10, 23, 29-31, 49].

#### *3.2.1 Extracting QoIs from the flow computations for two different systems of interest*

*System 1: Gas-particle mixtures*

The first system of interest involves fluids interacting with solid particles, and has applications in dusty gas flows [78-80], supersonic combustion of droplets and solid particles [81] and gas-solid fluidized beds [82, 83]. In this system, the meso-scale morphology changes due to volume fraction, size, shape and spatial distribution of particles in the system. Process scale shock-particle interactions commonly contain large numbers of particles, and their spatial placement affects the overall drag and velocity fluctuations in the system [84]. While mesoscale simulations have been performed in previous works to understand the effects of volume fraction [20] and arrangement [85], the effects of other spatial morphological quantities at the macro level response of the system have not yet been evaluated. For typical particle systems the microstructure is stochastic, requiring quantification using statistical measures. Uncertainty quantification in the closure models [85] also requires the arrangement of particles as an independent parameter. A key contribution of this paper is to demonstrate the connection between mesoscale spatial particle arrangement and mesoscale QoIs.

The mesoscale QoIs considered for this study are the spatio-temporally averaged drag coefficient $\langle \overline{C_{\mathrm{D}}} \rangle$, pseudo-turbulent stress $\langle \tilde{S}_{\mathrm{ij}} \rangle$, and pseudo-turbulent kinetic energy $\langle \widehat{PTKE} \rangle$ [86]. The equations for calculating these quantities [86] are given in the appendix (Equations (A 14), (A 15), and (A 16)). Figure 1 shows the process of connecting the structure to mesoscale QoIs arising from the mesoscale physics. The levelset field obtained from the images (Figure 1 (a)) are used for particle resolved meso-scale, sharp interface computations (Figure 1 (d)) [23]. Figure 1 (d) illustrates the meso-scale QoIs extracted from the simulations in System 1.

*System 2: Pressed heterogeneous energetic (HE) materials*

The microstructure in HEs plays a crucial role in its response to shock loading conditions [29, 87, 88]. Crystal size and void distributions have been shown to affect the sensitivity of the samples under shock loading [89-91]. Experiments using different meso-structures of pressed HMX samples show morphology-dependent sensitivities [92, 93]. Samples with higher volume fractions of voids have shown higher sensitivity under shock loading [87]. Mesoscale hydrodynamic simulations of pore collapse in HMX have also revealed a strong dependence of post-shock hotspot temperature on the pore morphology [92, 93].

Therefore, the meso-structure of a HE material is strongly linked to its shock sensitivity i.e. initiation of chemical reaction and hotspot growth [94, 95]. In the following, we use pressed HMX images and a levelset-based structural quantification approach to establish S-P linkages.

S-P linkages are studied for two types of HEs. Sample A is a Class V pressed HMX material (Figure 12 (a)) and sample B is a Class III pressed HMX material (Figure 13 (a)) [1]. The SEM images of sample A and B show intragranular and inter-granular voids. The mesoscale computation methodology associated with the void collapse process and initiation of reaction in such imaged meso-structures are described in detail in previous works [17, 29, 96]. Here, mesoscale computations are performed to study the dynamics of void collapse that leads to hotspot formation, followed by chemical decomposition of HMX. Several mesoscale QoIs are used to study the effects of the structure on the mesoscale physics i.e. void collapse, hotspot initiation, and growth. The temperature $T(\boldsymbol{x},t)$, and product species mass fraction denoted by $Y_4(\boldsymbol{x},t)$, are field variables (see Equations (A 7), (A 17), (A 18)) that measure the intensity of the void collapse and the reaction progress in the resulting hotspot. The total mass fraction of the solid HMX converted to reaction product gaseous species in the entire domain is denoted by $F$ (Equation (A 18)). The rate of production of reaction products $\dot{F}$ is tracked as a QoI which quantifies the sensitivity of a sample of pressed HMX. Here we show how morphological features of voids are linked to the sensitivity QoI $\dot{F}$ .

First, in 3.3, we establish the S-P linkage for shocked gas-solid flows, where the embedded solids are regular in shape. Then, in Section 3.4, the structural metrics are combined with QoI extraction for the irregular shapes in the pressed HMX meso-structure, a case in which the image-derived microstructure is topologically complex.

## 3.3 System 1: Establishing structure-property linkages for shocked particle-laden gas systems

### *3.3.1 Quantification of structure: degree of obstruction in regularly shaped particle clusters*

In this section, the structure of particle clusters is characterized using the morphological metrics described in Section 2.1.4. The degree of obstruction is an important factor that determines the effect of particle arrangement on mesoscale QoIs such as $\langle\overline{C_{\mathrm{D}}}\rangle$, $\langle\widetilde{PTKE}\rangle$, and $\langle\tilde{S}_{22}\rangle$ [97]. To study the effects of particle arrangements, seven different arrangements –one inline, one staggered and five random arrangements – are used with a particle volume fraction $\phi = n\pi {d_{\mathrm{cyl}}}^2/4$ of 5%, impacted by a shock of $Ma = 3.5$. The particle diameter is kept constant at $d_{\mathrm{cyl}} = 0.001$. The arrangements are generated using MATLAB [76] in the form of black-and-white images. The images are then segmented [27] to obtain the levelset field shown in Figures 10 (a), (d), and (g).

The tortuosity in the x-direction $T_{\mathrm{x}}$ is calculated using Equation (27) as the average of the tortuosities $T_{\mathrm{xi}}$ for various paths numbered i. Figure 8 shows the $T_{\mathrm{xi}}$ corresponding to each path for seven different particle arrangements. The inline arrangement shown in Figure 10 (a) has $T_{\mathrm{xi}} \approx 1$ for all the paths; keeping all the other factors $(d_{\mathrm{cyl}}, \phi)$ constant the inline arrangement has the lowest $T_{\mathrm{x}}$. This implies that the degree of obstruction to the incoming shock is lower for the inline arrangement compared to other arrangements. For example, for the staggered arrangement shown in Figure 10 (d) $T_{\mathrm{xi}} \approx 1.4$ for all the paths. The five different random arrangements of particles have tortuosity values $1.2 < T_{\mathrm{xi}} < 1.35$ for the various paths. The averaged tortuosity $T_{\mathrm{x}}$ for all the randomly arranged particle systems shows that their degree of obstruction lies in between inline and staggered arrangements. Therefore, the tortuosity metric quantifies the degree of obstruction of a particular microstructure.

*3.3.2 The structure-property linkage for particle clusters in shocked flows*

The effect of particle loading (i.e. solid volume fraction $\phi_s$) on the drag and pseudo-turbulent stresses in a particle cloud has been studied in previous work [22, 32, 86]. To study the effects of the structure, i.e. particle arrangement, numerical simulations are performed to compute mesoscale QoIs: $\langle\overline{C_D}\rangle$, $\langle\widehat{PTKE}\rangle$, and $\langle\tilde{S}_{22}\rangle$. In Figure 10 (a), the inline arrangement of particles is shown at $t^* = 0$, where $t^* = \frac{t}{t_{\mathrm{ref}}}$ and reference time scale $t_{\mathrm{ref}} = \frac{l_{\mathrm{ref}}}{u_s}$, $l_{\mathrm{ref}}$ is the reference length scale and its value is 1.0 based on the dimensions of the square that contains the particle cluster and $u_s$, is the incident shock speed. At $t^* = 0.706$, the incident shock has traversed nearly half the length of the particle cluster. Each column of particles in Figure 10 (b) forms its own reflected shock-system which merge, forming a sequence of reflected shock structures. Baroclinic vortices are produced behind the particles and are advected primarily in the direction of shock propagation at this stage. The vortical structures interact with the shocks in the wake of each column of particles. Figure 10 (c) shows the Schlieren plot at $t^* = 2.06$ after the shock has traveled through the entire domain. At this late stage, the vortices are also observed to be advected away from the particle cluster in the direction transverse to shock propagation.

In Figure 10 (d) the numerical Schlieren for the staggered cluster of particles is shown at $t^* = 0$. As the shock travels through nearly half the length of the control volume in Figure 10 (e), vortical structures interact with each other and with a system of shocklets in the wake. Unlike in Figures 10 (a - c), in the inline arrangement, the sequence of reflected shocks from subsequent columns of particles are not observed in Figure 10 (e). Hence, reflected shocks from particles in a single column can interact constructively as well as destructively. Since the second column of particles is staggered with respect to the first, there is a greater obstruction to the primary shock compared to the inline case. This leads to the clustering of vortices in the initial few columns of the staggered arrangement when compared to the inline arrangement in Figure 10 (b). For the staggered arrangement, even at earlier times, e.g. at $t^* = 0.706$, vortical structures are observed to be advected transversely to the particle clusters. Furthermore, the curvature of the transmitted shock front after passing through the entire particle cluster, shown in Figure 10 (f), is higher for the staggered arrangement than for the inline arrangement (Figure 10 (c)). This higher curvature can be attributed to the above noted greater obstruction of the incident shock wave in the staggered configuration.

Figure 10 (g) shows a randomly arranged particle cluster. As expected, the vortical structures in Figure 10 (h) are asymmetric in this case. Figure 10 (i) shows complex transverse vortical patterns similar to the staggered arrangement. At $t^* = 2.06$, the transmitted shock front can be observed to have a higher curvature than in the inline case as it exits the particle cluster. Overall, the incident shock faces higher obstruction in the staggered and random arrangements compared to the inline arrangement. Higher obstruction to the shock passage leads to higher velocity fluctuations, i.e. vortical advection in the transverse direction. These aspects of the physics are captured by the QoIs calculated for each arrangement, as discussed in the following.

A comparison of the time evolution of the average drag coefficient $\overline{C_D}$ for the inline, staggered and five different random arrangements is shown in Figure 11 (a). The inline system of particles produces a higher initial peak drag in contrast with the other particle arrangements. This is because, for the inline arrangement, the shock faces more particles in the first column compared to the other arrangements. As the shock passes through the entire cluster of particles, the $\overline{C_D}$ for the inline system decays faster, i.e. the shock attenuates faster compared to all the other systems. Table 1 shows the $\overline{C_D}$ integrated from $t^*$ = 0 to 1, i.e. the time-averaged $\langle\overline{C_{\mathrm{D}}}\rangle$ as defined in Equation (A 12). The $\langle\overline{C_{\mathrm{D}}}\rangle$ is highest for staggered and lowest for inline arrangements of particles. As expected from Figure 11 (a), the $\langle\overline{C_{\mathrm{D}}}\rangle$ for the five randomly distributed particle systems lies between the inline and staggered arrangements.

Figure 11 (c) and (e) show the time evolution of $\widetilde{PTKE}$ and $\tilde{S}_{22}$ for different particle arrangements. It is observed that $\widetilde{PTKE}$ is higher for the staggered case than for random or inline arranged particles. Similarly, $\tilde{S}_{22}$ is also higher for the staggered arrangement compared to the inline or random arrangement of particles. Table 1 shows the time averaged velocity fluctuations $\langle\widetilde{PTKE}\rangle$, and $\langle\tilde{S}_{22}\rangle$. It can be seen that the staggered arrangement has the highest value of $\langle\widetilde{PTKE}\rangle$, and $\langle\tilde{S}_{22}\rangle$. On the other hand, inline arrangement has the lowest $\langle\widetilde{PTKE}\rangle$, and $\langle\tilde{S}_{22}\rangle$ among the seven different arrangements.

Figure 11 (b) shows the correlation between $\langle\overline{C_{\mathrm{D}}}\rangle$, the spatio-temporally averaged drag coefficient obtained in Equation (A 14) and the tortuosity $T_{\mathrm{X}}$. $T_{\mathrm{X}}$ varies from 1 to 1.4 depending on the particle arrangement. For the configurations with higher $T_{\mathrm{X}}$ the spatio-temporally averaged drag coefficient $\langle\overline{C_{\mathrm{D}}}\rangle$ is higher. This is expected because $T_{\mathrm{X}}$ quantifies the degree of obstruction faced by the incident shock. The staggered arrangement of particles shown in Figure 10 (d) has the highest $\langle\overline{C_{\mathrm{D}}}\rangle$ and $T_{\mathrm{X}}$. In contrast, the inline particle arrangement shown in Figure 10 (a) has the lowest $T_{\mathrm{X}}$ and therefore produces the lowest $\langle\overline{C_{\mathrm{D}}}\rangle$. All the randomly arranged particle configurations have $T_{\mathrm{X}}$ values between the inline and staggered arrangement.

Figure 10 (f) correlates $\langle\tilde{\mathrm{S}}_{22}\rangle$ with $T_{\mathrm{X}}$. $\langle\tilde{\mathrm{S}}_{22}\rangle$ is a measure of transverse velocity fluctuations in the flow field. Since the velocity fluctuations in the y-direction are also caused by the degree of obstruction, the $\langle\tilde{\mathrm{S}}_{22}\rangle$ increases linearly with increasing $T_{\mathrm{X}}$. Finally, in Figure 10 (d), the correlation between $\langle\widetilde{PTKE}\rangle$ and $T_{\mathrm{X}}$ is shown. Similar to the results seen with $\langle\overline{C_{\mathrm{D}}}\rangle$, $\langle\tilde{S}_{22}\rangle$, and $\langle\widetilde{PTKE}\rangle$ also directly correlates with the $T_{\mathrm{X}}$. Therefore, for shock-particle interactions, the structure-property linkage can be established using $T_{\mathrm{X}}$, and the mesoscale QoIs $\langle\overline{C_{\mathrm{D}}}\rangle$, $\langle\tilde{S}_{22}\rangle$, and $\langle\widetilde{PTKE}\rangle$. In summary, a useful structural metric linked with the resistance to shock passage in particle clusters is the tortuosity $T_{\mathrm{X}}$ which is quantified directly from binary images using the methods described in section 2.1.4.

### 3.4 System 2: S-P linkages for shock interaction with pressed HMX microstructures

We now study the relationship between micro-structure and initiation sensitivity of two different classes (labeled A and B) of pressed HMX materials. The void morphology and therefore the metrics used for characterization of these two samples are quite different. While sample A contains a large number of smaller but nearly uniformly sized voids, sample B contains a combination of various void sizes – with the larger voids exhibiting highly elongated branched structure. The following sections detail the void morphology characterization and mesoscale physical response for the two classes of HMX samples to establish S-P linkages.

#### *3.4.1 Quantification of structure: void morphological features for pressed HMX samples*

The methods developed in Section 2.1 are used to quantify the structure of the pressed energetic materials. Some of the important physical descriptors identified are void size $\Omega$, aspect ratio $\mathcal{A}$, orientation $\Theta$, interface to volume ratio $\ell$, curvature $K$, anisotropy $L_{\mathrm{S}}(\theta)$, nearest neighbor distance $\mathbb{D}$ and nearest interface distance $\mathbb{I}$. Each of these descriptors and their importance are discussed below.

*3.4.1.1 Void sizes and shapes*

[16] have shown that void size significantly affects rates of hotspot ignition and growth. The void area in 2-D $\Omega_{\mathrm{l}}$ for each void numbered l is calculated using Equation (2). A distribution of $\Omega_{\mathrm{l}}$ corresponding to the sample A in Figure 12 (a) is shown in Figure 12 (d). The mean $\Omega_{\mathrm{l}}$ is $0.0187\ \mu m^2$ with a standard deviation of $0.0301\ \mu m^2$. It can be also noted that there is a peak at $\Omega_{\mathrm{l}} = 0.01\ \mu m^2$, contributed by small voids having effective diameter in the range of $0.1\ \mu m$ spread throughout the domain (see Figure 12 (a)).

The orientation $\Theta_{\mathrm{l}}$ of voids is important to quantify because the intensity of void collapse generated hotspots depends significantly on this parameter [93]. Using Equation (11), orientations $\Theta_{\mathrm{l}}$ of individual voids numbered l are calculated and shown in Figure 12 (b). $\Theta_{\mathrm{l}}$ distribution in Figure 12 (e) shows a peak at $\Theta_{\mathrm{l}} = 0°$; this preferential orientation may reflect the directional aspect of the pressing force applied to make the HMX sample. The mean orientation direction of the voids is $28.30°$ with a standard deviation of $25.27°$. The high value of standard deviation indicates that the sample of HMX contains voids that have no predominant orientation.

Shape descriptors of voids such as the aspect ratio $\mathcal{A}_{\mathrm{l}}$ affect the physics of void collapse [18, 93]. Slender voids ($\mathcal{A} > 5$) have been shown to produce higher temperature hotspots due to a pinching mechanism[92, 93]. $\mathcal{A}_{\mathrm{l}}$ is quantified using Equation (13) and plotted for individual voids in Figure 12 (c). The voids in the HMX sample have $\mathcal{A}_{\mathrm{l}}$ ranging from 2 to 6. The long slender void numbered 1 has $\mathcal{A}_{\mathrm{l}} = 15$. The distribution of $\mathcal{A}_{\mathrm{l}}$ is shown in Figure 12 (f). The mean $\mathcal{A}_{\mathrm{l}}$ is 2.2 with a standard deviation of 5.2, implying most of the voids are elongated, with aspect ratio values $\mathcal{A}_{\mathrm{l}} = 2$.

The ratio of void perimeter to area $\ell_{\mathrm{l}}$ can impact the growth phase of hotspots as shown in previous work [18, 98]; hotspot shapes reflect the shapes of the voids from which they are formed. Therefore voids with higher $\ell_{\mathrm{l}}$ are more likely to produce hotspots with higher $\ell_{\mathrm{l}}$; hotspots with higher $\ell_{\mathrm{l}}$ grow faster and therefore lead to higher rates of chemical energy release.

Based on the average void size $\Omega_{\mathrm{l}}^{\mathrm{avg}}$ calculated from the $\Omega_{\mathrm{l}}$ distribution in Figure 12 (d), the average effective diameter of voids $d_{\mathrm{void}}^{\mathrm{avg}} = 0.1\ \mu m$, where $d_{\mathrm{void}}^{\mathrm{avg}} = \sqrt{\frac{4\times\Omega_{\mathrm{l}}^{\mathrm{avg}}}{\pi}}$ ; the $\ell_{\mathrm{l}}$ corresponding to the effective diameter value is $20\ \mu m^{-1}$. The distribution of the $\ell_{\mathrm{l}}$ calculated using Equation (4) is shown in Figure 12 (g). The $\ell_{\mathrm{l}}$ distribution shows that there are a large number of voids with $\ell_{\mathrm{l}} > 30\ \mu m^{-1}$ which implies that the $\ell_{\mathrm{l}}$ of most voids in the HMX sample are statistically higher than those of corresponding cylindrical voids.

The surface curvature of the voids $\mathrm{K}_{\mathrm{l}}$ is obtained using Equation (7) for the HMX sample (Figure 12 (a)). The distribution of $\mathrm{K}_{\mathrm{l}}$ in Figure 12 (j) shows that $\mathrm{K}_{\mathrm{l}}$ values range from 15 to 50. $\mathrm{K}_{\mathrm{l}}$ may have secondary effects on the collapse profiles which may in turn affect the overall sensitivity of energetic material samples.

*3.4.1.2 Microstructural Anisotropy or Preferential Orientation of Features*

Microstructural anisotropy seeks to quantify preferential directionality in the void field in the pressed energetic material; this is a global measure that is quantified using the star length ($L_{\mathrm{S}}(\theta)$) described in Section 2.1.3. Figure 12 (k) shows a polar plot of star length $L_{\mathrm{S}}(\theta)$ as a function of orientation $\theta$ corresponding to the HMX image in Figure 12 (a). The purple markers in the figure indicate the magnitude of $L_{\mathrm{S}}(\theta)$ at various $\theta$. An ellipse fit to the polar plot of $L_{\mathrm{S}}(\theta)$ is used to calculate the anisotropy tensor $\mathcal{M}$, anisotropy index $A_{\mathrm{index}}$, and the global orientation $\Psi$ using Equations (19), (20), and (21) respectively. $\Psi$

quantifies the predominant orientation of the voids, and $A_{\text{index}}$ gives information on the degree to which the voids are elongated in the preferred orientation. For the sample in Figure 12(a), $\Psi = 4.38°$ and $A_{\text{index}} = 2.6$, suggesting that the majority of the voids are stretched horizontally. Low values of $A_{\text{index}}$ indicates that the voids are either circular or has no preferred orientation; higher values of $A_{\text{index}}$ indicate the presence of a large number of voids with high $\mathcal{A}_{\text{l}}$ having similar orientations. Therefore, the direction of shock loading may influence the overall sensitivity of a pressed mesoscale HE sample with high $A_{\text{index}}$ values.

*3.4.1.3 Nearest neighbor and nearest interface distributions*

Void-void interactions have significant impact on hotspot intensity and are caused by blast waves that propagate outwards from locations of void collapse; the relative proximity of voids dictates whether the blast waves will affect the collapse process of neighboring voids [18]. The nearest neighbor distance $\mathbb{D}_{\text{l}}$ measures the proximity of voids in the microstructure (based on inter-centroidal distances). $\mathbb{D}_{\text{l}}$ is measured using Equation (23) for each void in the pressed HE in Figure 12 (a); the distribution of $\mathbb{D}_{\text{l}}$ is shown in Figure 12 (h). This distribution can be used to infer that no two voids in the HE sample are closer than $0.12\ \mu m$, since the lowest $\mathbb{D}_{\text{l}}$ value is $0.12\ \mu m$. The mean $\mathbb{D}_{\text{l}}$ is $0.322\ \mu m$ with a standard deviation of $0.19\ \mu m$ .The nearest interface distance $\mathbb{I}_{\text{l}}$ (Equation (25)) is also used to quantify the proximity of voids to each other and is similar to nearest neighbor distance $\mathbb{D}_{\text{l}}$; $\mathbb{I}_{\text{l}}$ uses the inter-interface distances rather than inter-centroidal distance used in $\mathbb{D}_{\text{l}}$ measurements. It can be seen from Figure 12 (i) that for most of the voids in the sample of HMX the nearest void interface $\mathbb{I}_{\text{l}} = 0.2\ \mu m$. The frequency peaks at the right side of the $\mathbb{I}_{\text{l}}$ distribution shows the presence of isolated voids in the sample. The mean $\mathbb{I}_{\text{l}}$ of the distribution is $0.195\ \mu m$ with standard deviation of $0.175\ \mu m$; a lower mean $\mathbb{I}_{\text{l}}$ implies higher probability of void interactions.

*3.4.1.4 Local structural metrics : $\omega_l$ and void orientation and $\alpha_l$*

Figure 13 (a) shows that sample B has crystals that are larger in size compared to sample A in Figure 12 (a). The presence of larger crystals lead to larger voids in the material; these large voids exhibit tortuous or branched structures. Some of the large voids are slender in some parts and thick in others, i.e. they exhibit varying aspect ratio within the same void. Also, these branched voids cannot be characterized by a single orientation measure. Therefore, instead of characterizing these voids as single objects, it is more meaningful to capture the local structural metrics at each point within the voids. Figure 13 (e) shows the joint probability distribution of the local aspect ratio $\alpha_{\text{l}}$ and local orientation $\omega_{\text{l}}$. The z-axis gives the probability of finding regions within the void with the specified $\alpha_{\text{l}}$ and $\omega_{\text{l}}$ values. The largest peak in the joint distribution is at local aspect ratio $\alpha_{\text{l}} = 6$ and local orientation $\omega_{\text{l}} = 80°$; this peak arises because of the presence of region 1 in the void, shown in Figures 13 (b) and (c). The second largest peak is for $\alpha_{\text{l}} = 6$ and $\omega_{\text{l}} = 30°$ is due to the lower part of the void, i.e. region 2 in Figures 13 (b) and (c). Therefore, the joint probability distribution functions of $\alpha_{\text{l}}$ and $\omega_{\text{l}}$ captures the local structure, orientation, and shape of the large branched voids.

*3.4.2 Structure-property linkage for pressed energetics*

To link the structure that was quantified in the previous section with mesoscale physics of shock interacting with pressed HEs, simulations are performed by applying a shock pressure of $9.5\ GPa$ and shock pulse duration of $1.132\ ns$. Sample A of size $10\ \mu m \times 5 \mu m$ is shown in Figure 12 (a), and sample B of size $6.22 \mu m\ \times\ 5.59 \mu m$ is shown in Figure 13 (a). Figure 14 (a) shows the simulation domain before the shock loading is applied, where voids are delineated by white regions and the blue regions represent solid HMX. Application of shock loading leads to void collapse which produces localized high temperature hotspots.

#### *3.4.2.1 Physics of hotspot formation and growth in imaged HMX meso-structure*

Three phases are observed during the void collapse process in response to shock loading for real HMX meso-structures:

Ignition Phase: The initial period immediately after the collapse of voids is the ignition phase [16, 94], leading to the formation of hotspots. Figure 14 (b) shows this initial period of shock passage through the void field, at $t = 0.97\ ns$. At this instant, the shock has collapsed nearly half of the voids in the domain. At the sites of void collapse local temperature values of $3000\ K$ are seen in Figure 14 (b); these hotspots arise from heating of solid HMX due to plastic work prior to void collapse, followed by much higher temperature rise due to material jetting and impact [99]. The hotspot temperatures are high enough to cause initiation of chemical reaction. At $t = 1.55\ ns$, as shown in Figure 14 (c), almost all the voids in the domain have collapsed. However, most of the hotspots are still in the ignition phase, i.e. chemical reactions at these sites are still being initiated. As time elapses, the hotspots may either diffuse away (sub-critical) or enter the steady growth phase (critical).

Growth Phase: During the growth phase shown in Figure 14 (d), a hotspot could either be super- or sub-critical [94, 100]. For the super-critical hotspots, exothermicity of the chemical reactions leads to the progression of a reaction front into the surrounding medium; for these classes of hotspots the temperature attained after initial collapse further increases. On the other hand, for sub-critical hotspots that are insensitive to shock loading, the hotspot is quenched by diffusion. The end of growth phase is marked by the super-critical hotspots beginning to merge with each other [3, 100].

Completion Phase: In the completion phase nearly 70% of the domain is engulfed with hotspots which continue to grow until all the solid HMX material is converted to gaseous products (Figure 14 (f)). Towards the end of the completion phase the chemical reaction rate begins to fall, since the reactant species i.e. solid HMX is consumed.

#### *3.4.2.2 Effect of void morphology on mesoscale physics*

In this section the void morphological features quantified in section 3.4.1 for the pressed HMX samples are linked to the hotspot physics calculated from the mesoscale computations. To visualize the effects of void morphology on hotspot characteristics, Figures 14 (a) to (e) show various stages of hotspot formation and growth. The void size is an important metric in determining the void collapse physics and resulting hotspot characteristics [99]. The void numbered $\mathrm{l} = 4$ is much smaller in size compared to $\mathrm{l} = 1$, or $\mathrm{l} = 2$, and produces a smaller hotspot as shown in Figure 14 (b). The resulting hotspot quenches due to thermal diffusion at $t = 2.42\ ns$. On the other hand, voids $\mathrm{l} = 1$, and $\mathrm{l} = 2$ produce larger hotspots as seen at $t = 1.55\ ns$, which grow to reach the completion phase of the entire sample (Figure 14 (e)).

The effects of void orientation $\Theta_\mathrm{l}$ are observed by comparing the collapse process of voids numbered $\mathrm{l} = 3$, and $\mathrm{l} = 6$. The $\Theta_\mathrm{l}$ plot in Figure 12 (b), shows $\Theta_3 = 80°$, and $\Theta_6 = 5°$ and since $\Theta_3 > \Theta_6$, void numbered $\mathrm{l} = 3$ forms a smaller hotspot which is observed to diffuse out at $t = 0.97\ ns$. However, the void numbered $\mathrm{l} = 6$ has nearly the same size as $\mathrm{l} = 3$ but has a lower $\Theta_\mathrm{l}$ and forms a larger hotspot. This is because, for voids with higher $\Theta_\mathrm{l}$, the lead shock during the collapse creates a strong rarefaction wave that produces a cooling effect [93].

The aspect ratio $\mathcal{A}_l$ has a significant impact on the collapse mechanism of voids [18, 93]. To study the effect of void aspect ratio $\mathcal{A}_l$ on the hotspot behavior, the temperature fields around voids numbered $l = 2$ and 5 (Figure 12 (c)) are compared in Figure 14 (c). Both voids 2 and 5 have similar $\Omega_l$ with aspect ratios $\mathcal{A}_2 = 7$ and $\mathcal{A}_5 = 2.5$ respectively. It is seen that void 2 produces a much larger hotspot, in agreement with previous studies [18, 93]. Therefore, the structural metrics $V_l$, $\Theta_l$, and $\mathcal{A}_l$ influence the post void collapse hotspot formation and growth mechanisms.

#### *3.4.2.3 Effect of nearest neighbor and nearest interface of voids on mesoscale physics*

Aside from void morphological features, void-void interactions play an important role in hotspot formation and growth [18]. The probability of void-void interactions depends on the spatial arrangement of voids in a microstructure. The nearest neighbor $\mathbb{D}_l$ and nearest interface $\mathbb{I}_l$ distributions described in Section 2.1.4 are used to quantify the spatial arrangement of voids in samples A and B.

Void-void interactions occur when the collapse of a void produces reactive blast waves [18] that can affect the collapse of neighboring voids. The Schlieren image in Figure 15 (a) shows the incident shock wave collapsing the voids in the left part of the sample A leading to formation of reactive blast waves. The pressure contours in Figure 15 (b) show the high pressure regions created by these blast wave fronts. While the pressure behind the applied shock front is 9.5 $GPa$, the blast wave causes the local pressures to rise to 13 $GPa$.

The local regions marked by white boxes in Figure 15 (a) and (b) are used to describe the behavior of reactive blast waves. Region 1 with a blast wave front of smaller radius has a pressure of 13 $GPa$, while region 2 with a larger blast front radius has a significantly lower pressure of 10.6 $GPa$. This observation indicates that the blast waves that emanate from the void collapse location weaken rapidly as they spread outward. The blast waves can also merge with the lead shock front and form a Mach stem (see Region 3 in Figure 15 (d)) [101]. Similar to reactive blast fronts, the pressure at the Mach stem locations reach values higher than that of the initial shock loading pressure. Therefore, post-collapse blast waves and Mach stems can affect the collapse process of the voids in the downstream regions of the sample. Figure 15 (c) shows several individual blast waves affecting the collapse process of the voids in the vicinity. In summary, the distance between voids or the proximity of void interfaces can affect the hotspot formation and growth mechanisms (i.e. void-void interactions), and metrics such as $\mathbb{D}_l$, $\mathbb{I}_l$ distributions can be used to capture the effect of such interactions.

#### *3.4.2.4 Complex void shapes: Local structural metrics and sensitivity*

Void aspect ratio $\mathcal{A}_l$ and orientation have significant effects on the hotspot characteristics [18, 93]. However, for complex shaped voids a single value of aspect ratio or orientation may not be sufficient to capture the physics of hotspot formation and therefore of sensitivity to shock load [102]. For example, in sample B shown in Figure 13 (a), the largest void may have local regions with varying sensitivity to shock loading. The void is elongated and consists of two distinct segments connected at a branch; the orientation of the different segments of the void with respect to the shock will be different as will the corresponding sensitivity. Overall void aspect ratio or orientation therefore may not be sufficient to characterize the sensitivity for such complex voids. Local metrics may need to be used, pertaining to different segments of the void. For example, for analysis purposes, the void in Figure 13 (a) is subdivided into two regions as indicated by the boxes: 1) vertically-oriented region, and 2) slanted region. To study the effects of local structural metrics on its sensitivity to shock loading, mesoscale reactive computations are performed using sample B. A shock pressure of 9.5 $GPa$ is applied from the left end of the domain shown in Figure 16 (a).

The temperature contours in Figure 16 can be correlated to the local morphological features $\omega_l(\boldsymbol{x},t)$ and $\alpha_l(\boldsymbol{x})$, which are calculated using Eqn. (10) and (12) respectively. The local metrics $\omega_l(\boldsymbol{x},t)$ and $\alpha_l(\boldsymbol{x})$ in Figure 13 (b) and (c) respectively show that the region 1 has $\omega_l(\boldsymbol{x},t) = 80°$ and $\alpha_l(\boldsymbol{x}) = 8$. The temperature rise caused by the collapse of region 1 is 800 $K$; this is a weak hotspot and is quenched by thermal diffusion. However, region 2 for which $\theta_l = 35$ ° and $\alpha_l(\boldsymbol{x}) = 8$ leads to a temperature rise of 3400 $K$ and this region proceeds to criticality. Therefore the hotspot formation mechanisms are strongly affected by the local structural metrics $\omega_l(\boldsymbol{x},t)$ and $\alpha_l(\boldsymbol{x})$ at different regions within a single void.

Local metrics are quantified using the methods described in 2.1.2 for eight different samples shown in Figure 17. To correlate the structural metrics with the criticality of the samples, reactive mesoscale computations are performed on the microstructural images using methods described in previous work [29]. A constant shock loading pressure of 5.4 $GPa$ is used for all the samples. The criticality of the samples is determined based on the time evolution of the meso-scale QoI, the reaction product fraction $F$, which is calculated using Equation (A 18). The results show that some samples have a higher sensitivity to shock loading than others, i.e. upon void collapse they readily produce a number of critical hotspots with faster growth rates. Therefore the sub- and super-critical samples are identified based on $F$ vs $t$ curves plotted for all the eight samples. Samples numbered 1, 3, 4, 5, 6, and 8 are identified as super-critical since the $F$ increases monotonically with $t$. Higher slopes of the $F$ vs $t$ curves ($\dot{F}$) indicate higher sensitivity of the samples; sample 8 shows the highest $\dot{F}$ among all the eight samples. The $F$ for the sub-critical cases had values less than 0.1 for times up to $t = 60\ ns$; sample numbers 2 and 7 are identified as sub-critical.

For each of the eight samples, $\omega(\boldsymbol{x})$ and $\alpha(\boldsymbol{x})$ are obtained as probability density distributions and shown in Figure 13 (d) and (e) respectively. These probability density distributions are used to correlate the sensitive regions within the samples to their criticality. First, the volume occupied by sensitive regions within the samples is computed. The sensitive regions are defined based on a previous study where it was shown that voids oriented at angles $0° - 40°$ are more likely to ignite and produce hotspots with higher growth rates [18]. Therefore, $\omega^{\text{sensitive}}$, which distinguishes the regions in the $\omega(\boldsymbol{x})$ space with high sensitivity is considered to be in the range $0° < \omega(\boldsymbol{x}) < 40°$ (Figure 13 (c)). $\omega^{\text{sensitive}}$ is obtained from:

$$\omega^{\text{sensitive}} = \frac{1}{\Omega}\int_{\Omega} \zeta^{\omega}(\boldsymbol{x})\, d\Omega \text{ where } \zeta^{\omega}(\boldsymbol{x}) = \begin{cases} 1 & if\ 0° < \omega(\boldsymbol{x}) < 40° \\ 0 & \end{cases} \tag{29}$$

where $\zeta^{\omega}(\boldsymbol{x})$ is a sensitivity indicator function. Similar to $\omega(\boldsymbol{x})$, it has been observed in a previous study that voids with aspect ratio greater than 3 have a higher probability of going critical [18]. Therefore, an indicator function $\zeta^{\alpha}(\boldsymbol{x})$ is defined such that it has a value of 1 only in the regions with $\alpha(\boldsymbol{x}) > 3$. $\alpha^{\text{sensitive}}$ is then defined by the following equation:

$$\alpha^{\text{sensitive}} = \frac{1}{\Omega}\int_{\Omega} \zeta^{\alpha}(\boldsymbol{x})\, d\Omega \text{ where } \zeta^{\alpha}(\boldsymbol{x}) = \begin{cases} 1 & if\ \alpha(\boldsymbol{x}) > 3 \\ 0 & \end{cases} \tag{30}$$

The criticality of the voids obtained from the mesoscale computations for the eight different pressed HMX samples are linked to the quantities $\omega^{\text{sensitive}}$ and $\alpha^{\text{sensitive}}$ using a support vector machine (SVM) [103]. The question being asked is: can the SVM predict the criticality of the samples based on the $\omega^{\text{sensitive}}$ and $\alpha^{\text{sensitive}}$ values of that sample for a given loading condition? Therefore, the SVM is trained using $\omega^{\text{sensitive}}$ and $\alpha^{\text{sensitive}}$ as predictors to identify the super- and sub-critical samples.

The result of the classification using SVM is shown in Figure 18 (b) where the $x$-axis denotes $\omega^{\text{sensitive}}$ and $y$-axis denotes $\alpha^{\text{sensitive}}$ and the black markers correspond to each sample. The red and blue colors in the plot are the regions predicted by the SVM to be super-critical and sub-critical respectively in the $\omega^{\text{sensitive}}$ and $\alpha^{\text{sensitive}}$ plane. It is observed that the samples with lower values of $\omega^{\text{sensitive}}$ and $\alpha^{\text{sensitive}}$ lie in the bottom left corner and are sub-critical (samples 2 and 7). Similarly, the samples with higher $\omega^{\text{sensitive}}$ and $\alpha^{\text{sensitive}}$ lie in the top right part of the plot (samples 1,3,4, 5, 6, and 8) and are super-critical.

Based on the above classification through S-P linkages, quantities such as $\omega^{\text{sensitive}}$ and $\alpha^{\text{sensitive}}$ can make predictions regarding the sensitivity of heterogeneous energetic material samples with complicated void structures. The quantities $\omega^{\text{sensitive}}$ and $\alpha^{\text{sensitive}}$ can be extracted using the levelset-based structural metrics developed in this work. Using such metrics to connect structure and properties, sensitivity of an imaged sample can be assessed in a machine learning framework using algorithms like the geometric convolutional neural networks (G-CNNs [104, 105]). This, along with establishing structure-property-performance linkages for imaged energetic material microstructures, is the subject of ongoing work.

## 4 CONCLUSIONS

A framework is presented for establishing structure-property (S-P) linkages for heterogeneous materials subject to shock loading which can be used for multi-scale modelling of such materials. The methodology for levelset-based structure quantification is general and can be used for other engineering applications requiring microstructure-aware predictive models. Here, the framework is demonstrated in the specific context of two shocked multi-material flows: 1) The dynamic of particle clouds under shock loading, pertaining to blast wave propagation or multiphase flows in supersonic combustion engines, and 2) the response of heterogenous energetic materials to shock insults.

The structural metrics are quantified from microstructure images by first converting them into levelset fields [27]. Using the levelset field, several structural metrics including orientation ($\Theta$), aspect ratio ($\mathcal{A}$), anisotropy using star length ($L_{\text{S}}$), tortuosity ($T_{\text{x}}, T_{\text{y}}$), and other spatial pattern metrics are developed and their linkage to the mesoscale physics are established.

The structural metrics are linked to the properties (i.e. homogenized mesoscale QoIs). The QoIs are calculated in both the applications using resolved meso-scale simulations. In the context of particle laden gases it is shown that the tortuosity ($T_{\text{x}}$) correlates directly to the mesoscale QoIs, viz. the spatio-temporally averaged drag coefficient $\langle\overline{C_{\text{D}}}\rangle$, pseudo-turbulent kinetic energy $\langle\widetilde{PTKE}\rangle$, and pseudo-turbulent stress tensor $\langle\tilde{S}_{22}\rangle$. For shocked heterogeneous energetic materials, structural metrics are used to quantify void morphology via size $\Omega$, interface perimeter to area ratio $\ell$, surface curvature $\mathrm{K}$, orientation $\Theta$, aspect ratio $\mathcal{A}$, nearest neighbor distance $\mathbb{D}$, and nearest interface distance $\mathbb{I}$. These morphological metrics are shown to affect the post collapse hotspot characteristics (size, temperature) which in turn significantly affect the sensitivity of samples.

For microstructures containing topologically complex voids, local structural metrics such as local aspect ratio $\alpha(\boldsymbol{x}, t)$, and orientation $\omega(\boldsymbol{x}, t)$ are used to isolate regions of sensitivity in the heterogenous energetic material samples. The extension of this ability to relate local metrics of microstructure to the computed response at the meso-scale is to employ images directly in a geometric machine learning algorithm. A preliminary S-P linkage in this context is demonstrated, using support vector machines to classify imaged micro-structure samples as sub- and super-critical.

This framework can be extended to three-dimensional systems where S-P linkage is necessary for predictive material modelling. We are also engaged in efforts to establish the structure-property-performance linkage for heterogenous energetic materials, where the performance is quantified by performing meso-informed macroscale simulations in a multi-scale framework [17].

## ACKNOWLEDGEMENTS

The authors gratefully acknowledge the financial support from the Air Force Research Laboratory Munitions Directorate (AFRL/RWML), Eglin AFB, under contract number FA8651-16-1-0005 (Program Manager: Dr. Angela Diggs).

# APPENDIX

## A1. Governing equations and numerical method for Meso-Scale Computations

The hyperbolic conservation laws for mass, momentum and energy are solved:

$$\frac{\partial \rho}{\partial t} + \frac{\partial(\rho u_{\mathrm{i}})}{\partial x_{\mathrm{i}}} = 0$$

$$\frac{\partial(\rho u_{\mathrm{i}})}{\partial t} + \frac{\partial\left(\rho u_{\mathrm{i}} u_{\mathrm{j}} + \sigma_{\mathrm{ij}}\right)}{\partial x_{\mathrm{j}}} = 0 \qquad \text{(A 1)}$$

$$\frac{\partial(\rho E)}{\partial t} + \frac{\partial\left(\rho u_{\mathrm{i}} E + \sigma_{\mathrm{ij}} u_{\mathrm{i}}\right)}{\partial x_{\mathrm{i}}} = 0$$

Here the density and velocity components of the material are denoted by $\rho$ and $u_{\mathrm{i}}$ respectively. The total specific energy is $E = e + \frac{1}{2} u_{\mathrm{i}} u_{\mathrm{i}}$, where $e$ is the specific internal energy. $\sigma_{\mathrm{ij}}$ is the Cauchy stress tensor and is given differently for the particle laden gas flow computations and pressed energetic materials.

The stress tensor $\sigma_{\mathrm{ij}}$ is:

$$\sigma_{\mathrm{ij}} = \Sigma_{\mathrm{ij}} - p\delta_{\mathrm{ij}} \qquad \text{(A 2)}$$

where $p$ is the pressure and $\Sigma_{\mathrm{ij}}$ refers to the deviatoric components of the stress tensor. For the gas, the deviatoric components $\Sigma_{\mathrm{ij}}$ of the stress tensor is zero and an ideal gas equation of state is used to calculate the pressure:

$$p = \rho e(\gamma - 1) \qquad \text{(A 3)}$$

$\gamma$ is the specific heat ratio with a value of 1.4. For the solid energetic materials, the deviatoric stress tensor $\Sigma_{\mathrm{ij}}$ is calculated by solving:

$$\frac{\partial\left(\rho \Sigma_{\mathrm{ij}}\right)}{\partial t} + \frac{\partial\left(\rho u_{\mathrm{i}} \Sigma_{\mathrm{ij}}\right)}{\partial x_{\mathrm{i}}} + \frac{2}{3}\rho G tr\left(D_{\mathrm{ij}}\delta_{\mathrm{ij}}\right) - 2\rho G D_{\mathrm{ij}} = 0 \qquad \text{(A 4)}$$

Where, $D$ is the strain rate tensor, and $G$ is the shear modulus. Pressure is obtained from the Birch-Murnaghan Equation of state [106], which can be written in the general Mie-Gruneisen form as:

$$p(\rho, e) = p_{\mathrm{k}}(\rho) + \rho \Gamma_{\mathrm{s}}[e - e_{\mathrm{k}}(\rho)], \qquad \text{(A 5)}$$

where:

$$p_{\mathrm{k}}(\rho) = \frac{3}{2} K_{T0}\left[\left(\frac{\rho}{\rho_0}\right)^{-7/3} - \left(\frac{\rho}{\rho_0}\right)^{-5/3}\right]\left[1 + \frac{3}{4}(K_{T0}' - 4)\left[\left(\frac{\rho}{\rho_0}\right)^{-2/3}\right] - 1\right], \qquad \text{(A 6)}$$

The Birch-Murnaghan equation of state is solved to obtain the dilatational response, and the deviatoric response of HMX is obtained by modeling perfectly plastic response under shock loading [106]. Thermal softening caused due to melting of HMX under shock loading is modeled using the Kraut-Kennedy relation [106]. The specific heat as a function of temperature is obtained from [106]. A three-step chemical decomposition model is applied for the chemical reaction of HMX [3]. Species formed by the chemical decomposition of HMX are evolved in time using the following species conservation equation:

$$\frac{\partial(\rho[Y_\mathrm{k}])}{\partial t} + \frac{\partial(\rho u_\mathrm{i}[Y_\mathrm{k}])}{\partial x_\mathrm{i}} = \dot{Y}_\mathrm{k} \tag{A 7}$$

where, $Y_\mathrm{k}$ corresponds to the mass fraction and $\dot{Y}_\mathrm{k}$ the mass production rate of the $\mathrm{k^{th}}$ species.

Spatial discretization of the governing equations is performed on a fixed Cartesian grid using a 3rd-order Essentially Non-Oscillatory (ENO) scheme [107]. A 3rd order Runge-Kutta scheme is used for explicit time marching. The computational domain contains objects in the form of particles in the case of gas-particle flows and voids in the case of pressed energetic materials. All interfaces between materials are delineated by a signed distance function or the levelset function $\varphi(\boldsymbol{x}, t)$. The levelset $\varphi(\boldsymbol{x}, t) < 0$ for regions in the interior of the objects and $\varphi(\boldsymbol{x}, t) > 0$ in the exterior and is defined in a narrow band for each interface [26]. If there are a total of $N_\mathrm{obj}$ objects in the domain, the levelset field for the $l^\mathrm{th}$ object is given by a separate narrowband levelset field $\varphi_\mathrm{l}(\boldsymbol{x}, t)$. No-penetration boundary conditions are applied at interfaces between the gas and solid using the modified ghost fluid method (GFM) [108]; appropriate boundary conditions on free surfaces and crystal-crystal contact conditions for the pressed HMX problem are prescribed [109]. A detailed explanation of the numerical framework used for the mesoscale simulations can be found in previous works [10, 23, 29-31, 49].

## A2. Spatial and temporal homogenization of QoIs for gas particle flow

The spatially-averaged drag, $\overline{F_\mathrm{D}}(t^*)$ is the mean drag experienced by all the particles in the domain. Details of the averaging methods can be found in previous work [86]. Here, $t^* = \frac{t}{t_\mathrm{ref}}$ and reference time scale $t_\mathrm{ref} = \frac{l_\mathrm{ref}}{u_\mathrm{s}}$. $l_\mathrm{ref}$ is the reference length scale and its value is 1.0 based on the dimensions of the square that contains the particle cluster. $u_\mathrm{s}$, is the incident shock speed. The spatially-averaged drag coefficient, $\overline{C_\mathrm{D}}(t^*)$ in the cluster is:

$$\overline{C_\mathrm{D}}(t^*) = \frac{\overline{F_\mathrm{D}}(t^*)}{\frac{1}{2}\rho_\mathrm{s} {u_\mathrm{s}}^2 d_\mathrm{eq}} \tag{A 8}$$

$\rho_\mathrm{s}$ is the density of static unshocked fluid, $u_\mathrm{s}$ is the shock speed and $d_\mathrm{eq}$ is the equivalent diameter of the particle cluster $d_\mathrm{eq} = \sqrt{\frac{4}{\pi}\Omega\,\phi}$ in domain of area Ω and volume fraction ϕ. The velocity fluctuations cause stresses in the gas phase, which appear as the pseudo-turbulent stress tensor, $\mathrm{S_{ij}}$ given by:

$$S_\mathrm{ij}(\boldsymbol{x}, t^*) = \frac{u_\mathrm{i}'(\boldsymbol{x}, t^*) u_\mathrm{j}'(\boldsymbol{x}, t^*)}{{u_\mathrm{s}}^2} \tag{A 9}$$

where $u_\mathrm{i}'(\boldsymbol{x}, t^*)$, is the velocity fluctuation field:

$$u_\mathrm{i}'(\boldsymbol{x}, t^*) = u_\mathrm{i}(\boldsymbol{x}, t^*) - \widetilde{u_\mathrm{g}}(t^*) \tag{A 10}$$

$\widetilde{u_g}(t^*)$, the gas slip velocity vector is calculated as follows:

$$\widetilde{u_g}(t^*) = \frac{\int_\Omega \rho u_l d\Omega}{\int_\Omega \rho d\Omega} \quad \text{(A 11)}$$

where, $\rho$ and $u_g$ are the fluid phase density and velocity component respectively.

The spatially-averaged pseudo-turbulent stress tensor, $\tilde{S}_{ij}(t^*)$ is:

$$\tilde{S}_{ij}(t^*) = \frac{\int_\Omega \rho S_{ij} d\Omega}{\int_\Omega \rho d\Omega} \quad \text{(A 12)}$$

The spatially-averaged pseudo-turbulent kinetic energy (PTKE [86]), i.e. $\widetilde{PTKE}(t^*)$ is:

$$\widetilde{PTKE}(t^*) \equiv \tilde{S}_{ii}(t^*) \quad \text{(A 13)}$$

The temporal homogenization is performed over a time $T^*$ beginning with the shock entering the cluster until the entire incident shock wave leaves the particle cluster. The spatio-temporally averaged drag coefficient $\langle\overline{C_D}\rangle$ is obtained as:

$$\langle\overline{C_D}\rangle = \frac{\int_{t^*=0}^{t^*=T^*} \overline{C_D}(t^*) dt^*}{\int_{t^*=0}^{t^*=T^*} dt^*} \quad \text{(A 14)}$$

The spatio-temporally averaged pseudo-turbulent stress tensor $\langle\tilde{S}_{ij}\rangle$ is given by:

$$\langle\tilde{S}_{ij}\rangle = \frac{\int_{t^*=0}^{t^*=T^*} \tilde{S}_{ij}(t^*) dt^*}{\int_{t^*=0}^{t^*=T^*} dt^*} \quad \text{(A 15)}$$

And the spatio-temporal average of the $PTKE$, $\langle\widetilde{PTKE}\rangle$ is defined as follows:

$$\langle\widetilde{PTKE}\rangle \equiv \langle\tilde{S}_{ii}\rangle \quad \text{(A 16)}$$

**A3. Spatial homogenization of QoIs in shocked HMX**

The product species mass fraction is denoted by $Y_4(x,t)$ and is evolved using the conservation equations for species (Equation (A 7)). The rate of chemical decomposition of HMX within the entire domain is calculated by accumulating the mass of the final gaseous products via:

$$M_{\text{reacted}}(t) = \int_\Omega \rho(\boldsymbol{x},t)\, Y_4(\boldsymbol{x},t)\, d\Omega \quad \text{(A 17)}$$

$M_{\text{reacted}}$ is calculated for the entire domain of area $\Omega$ and $\rho$ is the local density. The burned fraction of HMX, $F$, is the mass of the solid converted to reaction product gaseous species denoted by $F$ is useful to study the long term behavior for a field of voids. The fraction of fully burned HMX to the total mass of HMX in the control volume,

$$F(t) = \frac{M_{\text{reacted}}(t)}{M_{\text{HMX}}} \tag{A 18}$$

$M_{\text{HMX}}$ is the total mass of HMX present in the control volume $\Omega$ prior to the start of chemical reactions.

*Table 1: Mesoscale QoIs of shock particle interaction system obtained for various particle arrangements*

| **Arrangement** | $\langle\overline{C_D}\rangle$ | $\langle\widetilde{PTKE}\rangle$ | $\langle\tilde{S}_{22}\rangle$ |
|---|---|---|---|
| *Inline* | 0.178 | 0.094 | 0.029 |
| *Staggered* | 0.207 | 0.104 | 0.058 |
| *Random – 1* | 0.192 | 0.099 | 0.051 |
| *Random – 2* | 0.191 | 0.098 | 0.049 |
| *Random – 3* | 0.191 | 0.099 | 0.049 |
| *Random – 4* | 0.192 | 0.098 | 0.050 |
| *Random – 5* | 0.188 | 0.099 | 0.049 |

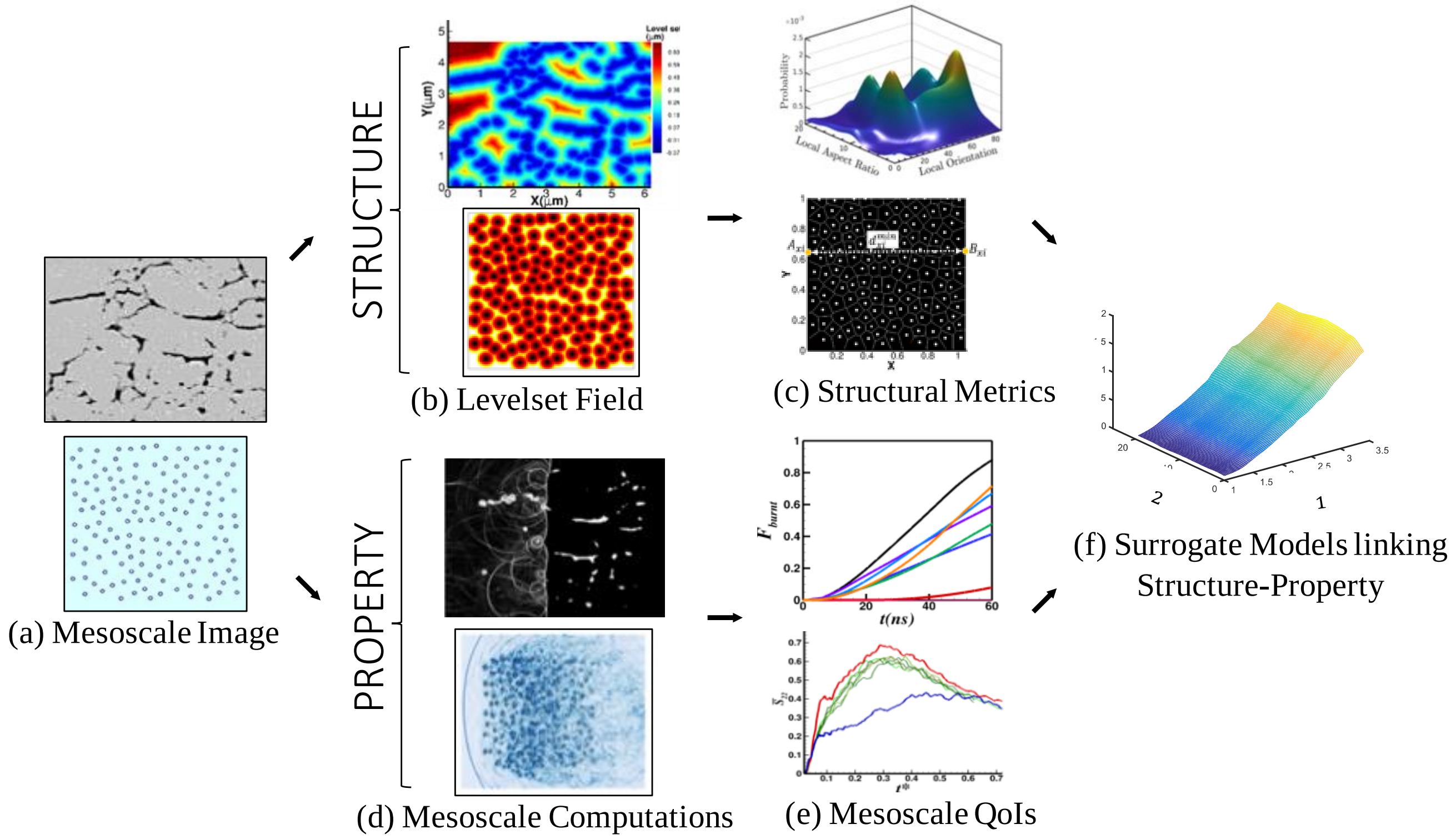

Figure 1: Schematic of the levelset based framework used for structure-property linkage. Shown are the pathways used for hierarchical multiscale modeling to bridge mesoscale to macroscale. (a) Images representing the structure of the materials at the mesoscale. (b) Levelset contour of the meso-structure obtained for both pressed HMX and particle clusters. (c) The structure at the mesoscale quantified using various morphological descriptors. (d) The properties required for macroscale simulations are derived from fully resolved mesoscale simulations. (e) Mesoscale QoIs are extracted from the mesoscale DNS. (f) The structure-property linkages are established using surrogate modelling of mesoscale QoIs as a function of morphological quantities $\psi_1$ and $\psi_2$.

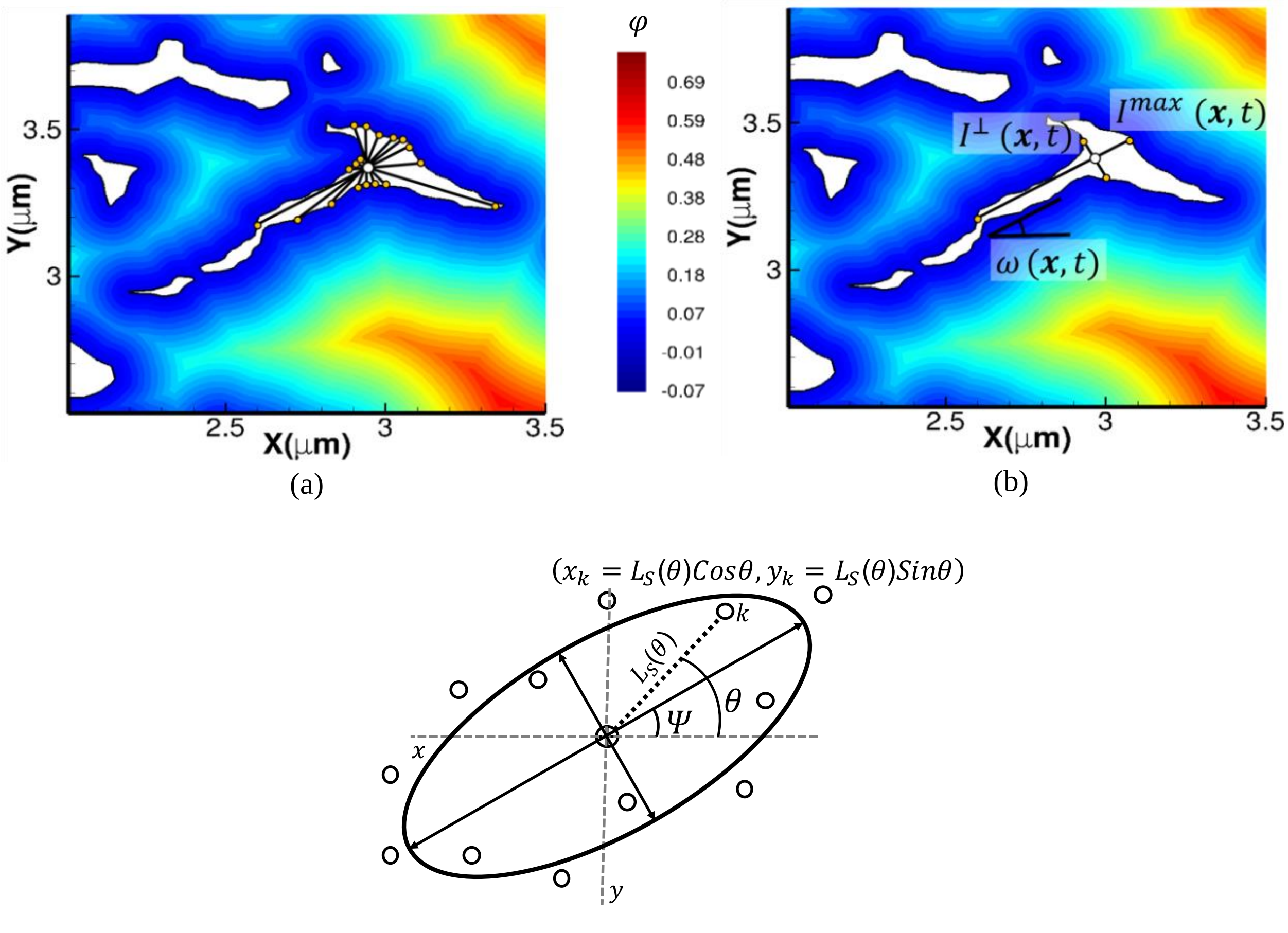

(c)

Figure 2: Illustration of the star length method applied to voids in pressed HE. (a) Line intercepts sampled towards multiple orientations $\theta$ from location $\boldsymbol{x}$. Yellow points are the locations at which the line intercepts intersect with the zero levelset of the object numbered $l$. (b) Illustration of volume orientation method for quantifying local orientation $\omega(\boldsymbol{x})$ and local aspect ratio $\alpha(\boldsymbol{x})$ for Cartesian locations $\boldsymbol{x}$ within objects. The figure indicates longest line intercept $I^{max}(\boldsymbol{x})$, and $I^{\perp}(\boldsymbol{x})$ perpendicular to $I^{max}(\boldsymbol{x})$ sampled from location $\boldsymbol{x}$. $\omega(\boldsymbol{x})$ is the local orientation given by the orientation of $I^{max}(\boldsymbol{x})$. (c) Ellipse fit to points obtained from the polar plot of $L_S(\theta)$. Figure shows a point numbered $k$ used to obtain $x_k$, and $y_k$ as a function of $L_S(\theta)$. The ellipse fit to the collection of points obtained from the $L_S(\theta)$ is oriented at $\Psi$. The ratio of the major axis to the minor axis of the ellipse gives the anisotropy index $A_{index}$.

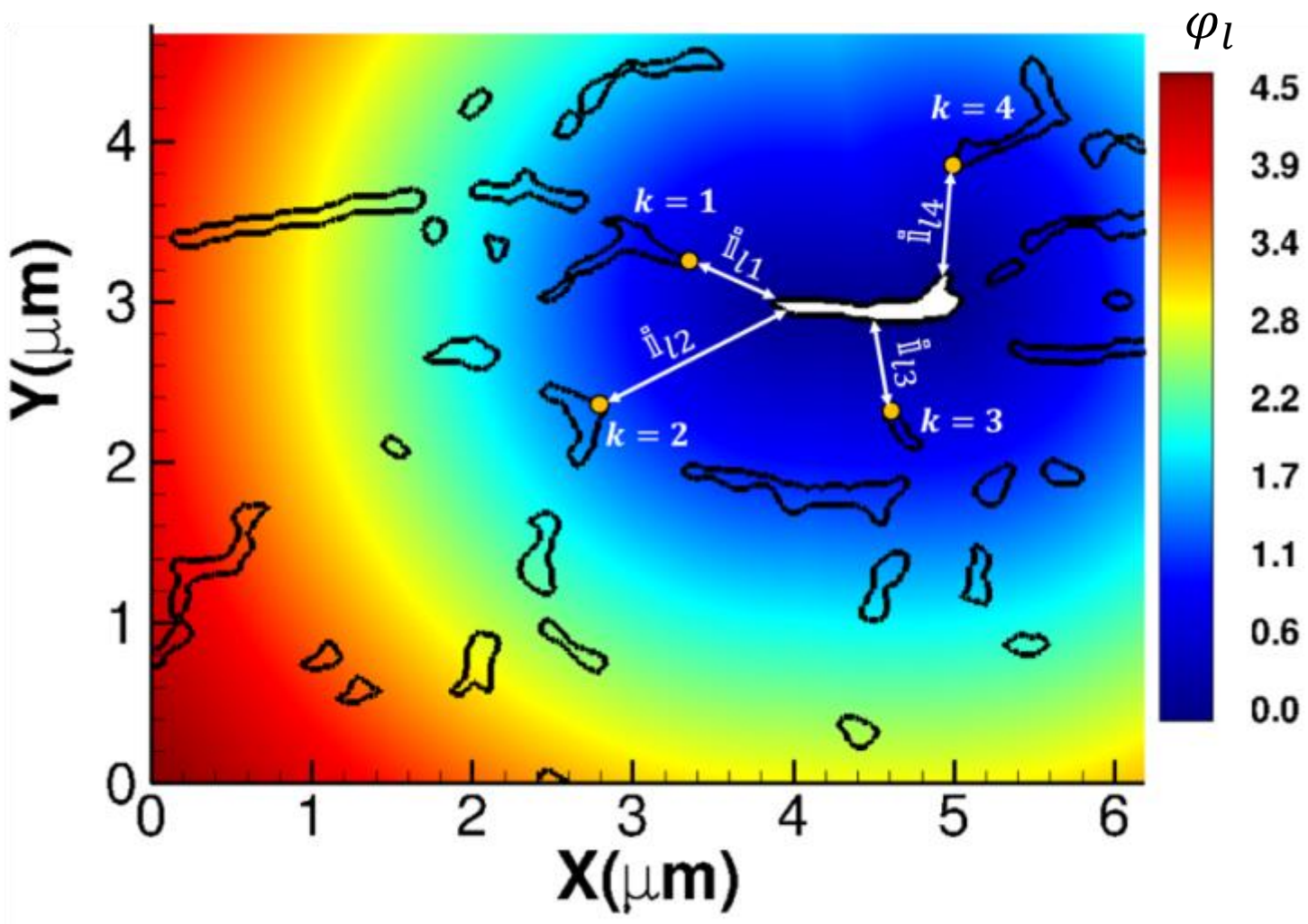

Figure 3: Illustration of calculating nearest interface distance from the levelset field. The levelset contour for object $l$ is shown above with the white region indicating negative levelset region. The interface or 0 levelset of the objects in the field are indicated by the black lines. The inter-interface distance $\mathbb{i}_{lk}$ between object $l$ and four other objects with $k$ values 1, 2, 3, 4 are also shown.

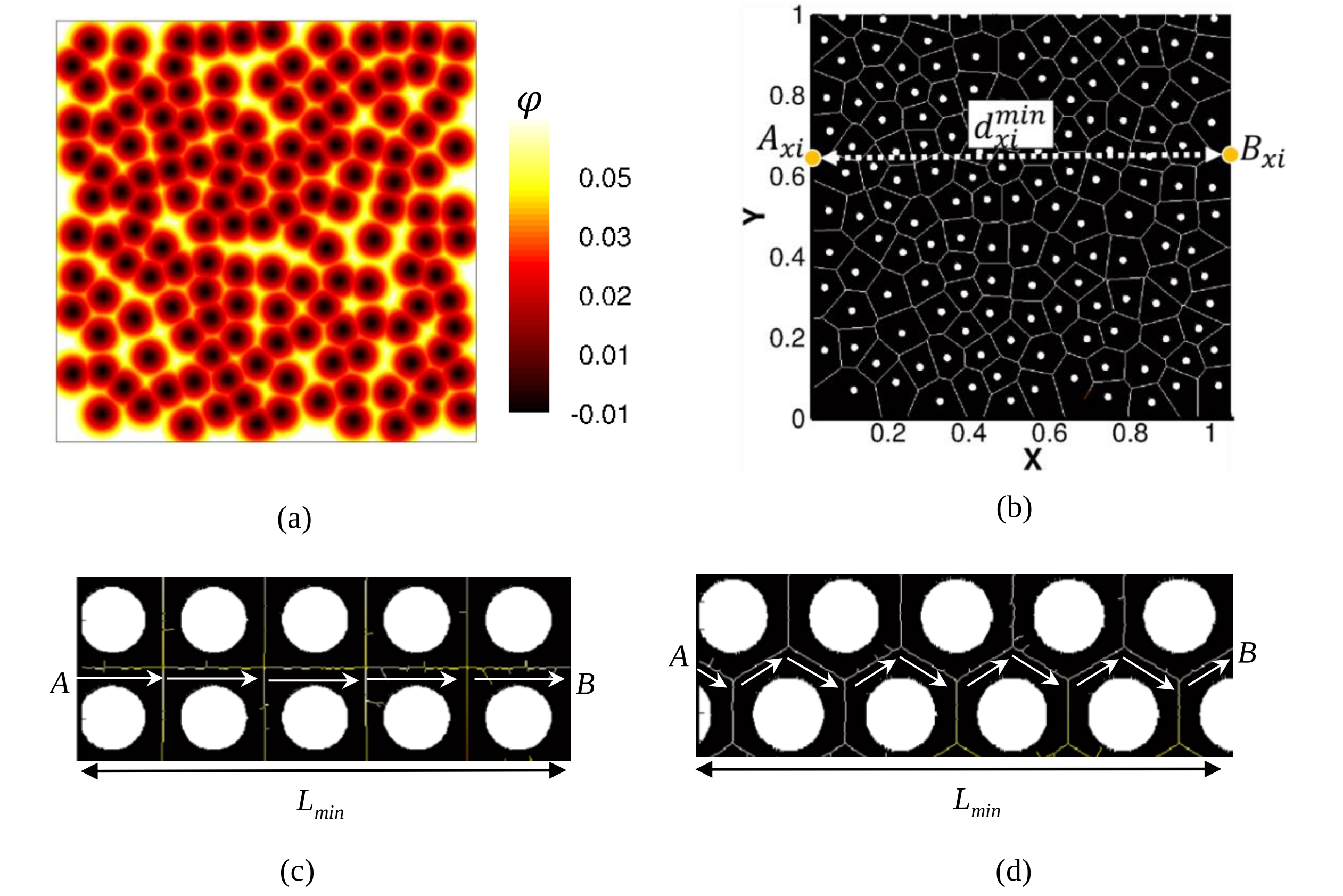

Figure 4: Illustration of method for calculating tortuosity. Medial skeletal pathways for two kinds of cylinder arrangement (a) levelset field $\varphi$ for a random cluster of particles. (b) Medial skeleton for the cluster of particles corresponding to the levelset field in (a). (c) represents inline arrangement of cylinders and (d) represents the staggered arrangement of cylinders. $L_{min}$ represents the straight line segments connecting points $A$ and $B$.

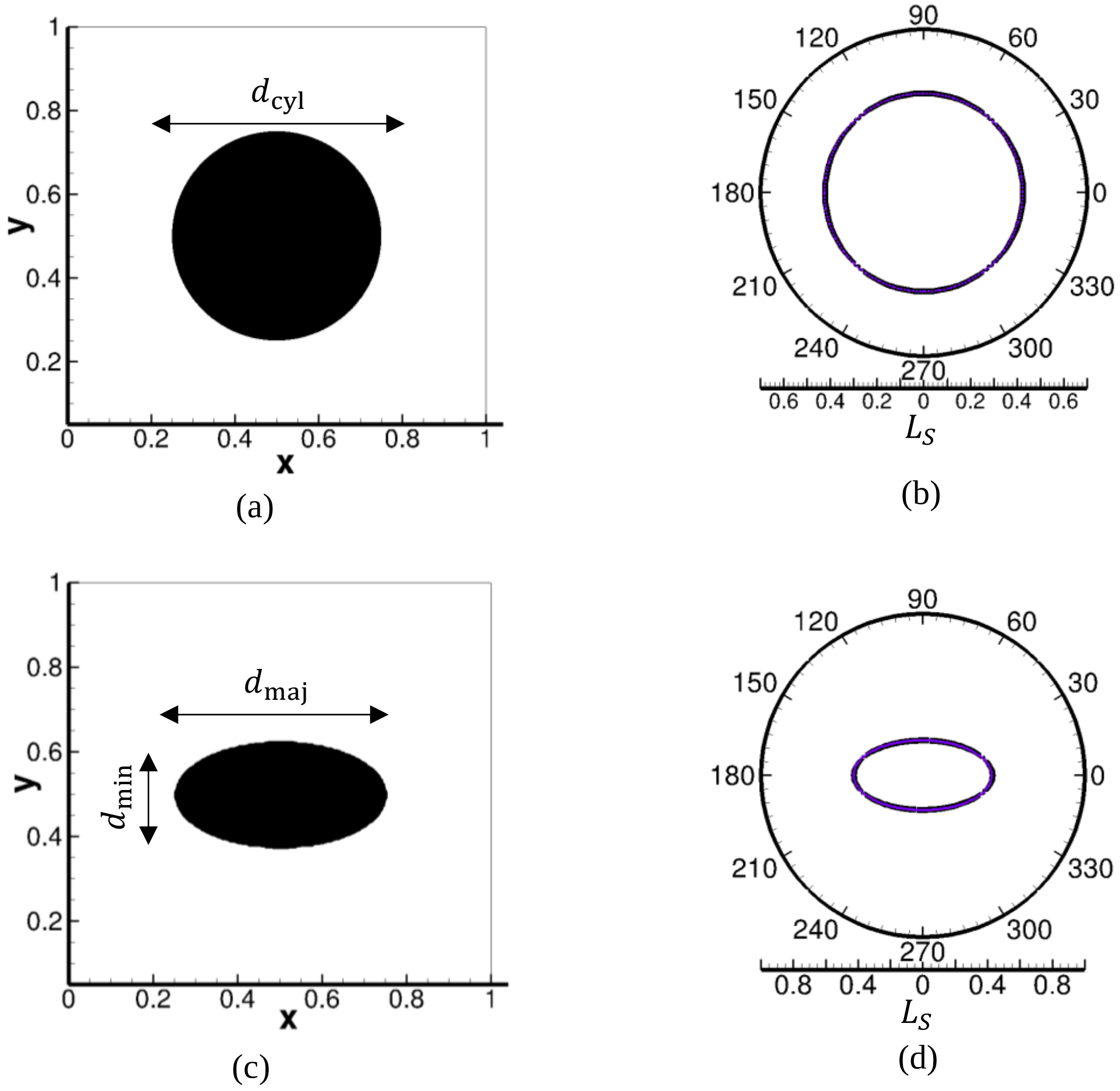

Figure 5: (a) A circle of radius 0.25 for which the star length is compared with a domain size of 1x1 and a grid resolution of $10^6$ points. (b) Polar plot of the $L_S(\theta)$ for the circle in (a) shown in purple and the ellipse fitted to the polar plot data is shown in black. (c) An ellipse with major axis length 0.505 and minor axis 0.255 in a 1x1 domain size and a grid resolution of $10^6$ points. (d) Polar plot of the $L_S(\theta)$ for an ellipse shown in (c).

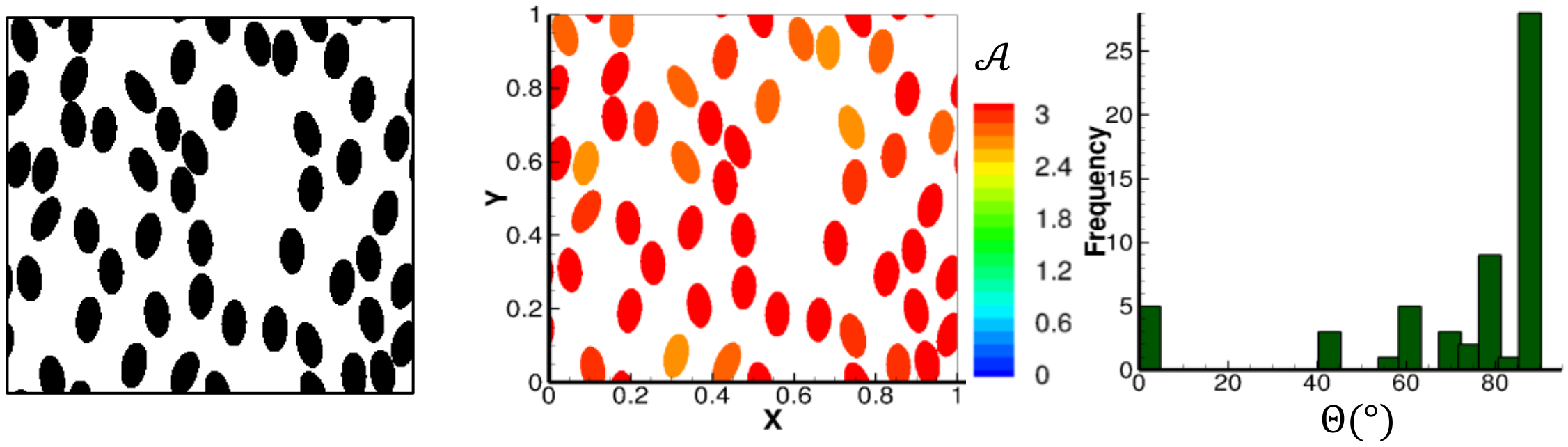

(a) (b) (c)

Figure 6: (a) Ellipse field with unit dimensions and ellipses with aspect ratio of 0.5 and volume fraction of 30%, the ellipses are oriented such that the mean orientation is 90° with a standard deviation of 10°. (b) aspect ratio $\mathcal{A}_l$ for each object (c) Object orientation angles $\Theta_l$ distribution plot.

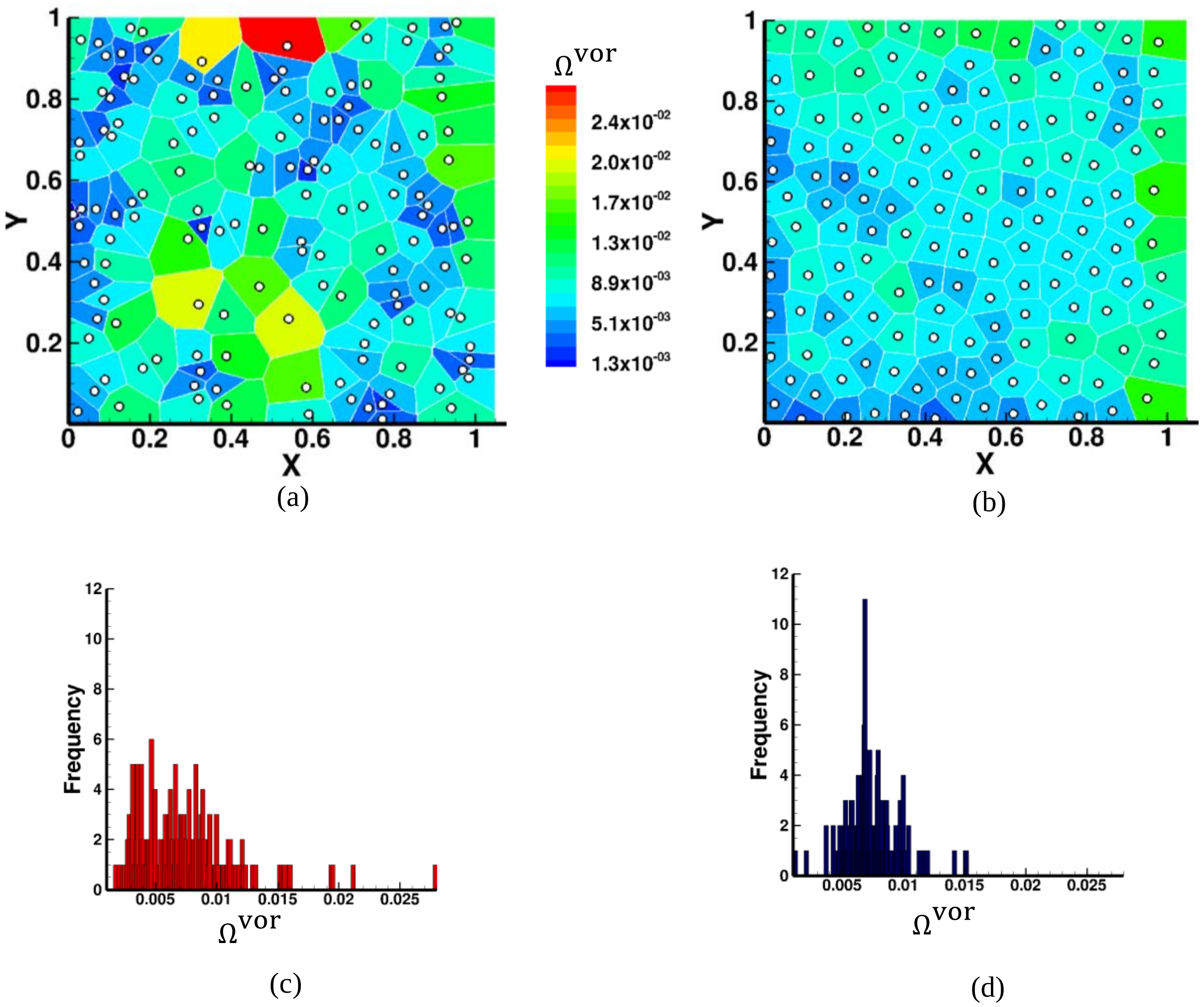

Figure 7: (a) and (b) Spatially distributed circular disks of diameter 0.020 with 10% volume fraction and a domain of size 1x1. (a) Contour plot showing the size of Voronoi cells $V_l^{vor}$ generated form a distribution of disks with the minimum inter-centroidal distance (a) 0.025 and (b) 0.045. (c) and (d) are the frequency distribution of $V_l^{vor}$ corresponding to the spatially distributed disks in (a) and (b) respectively.

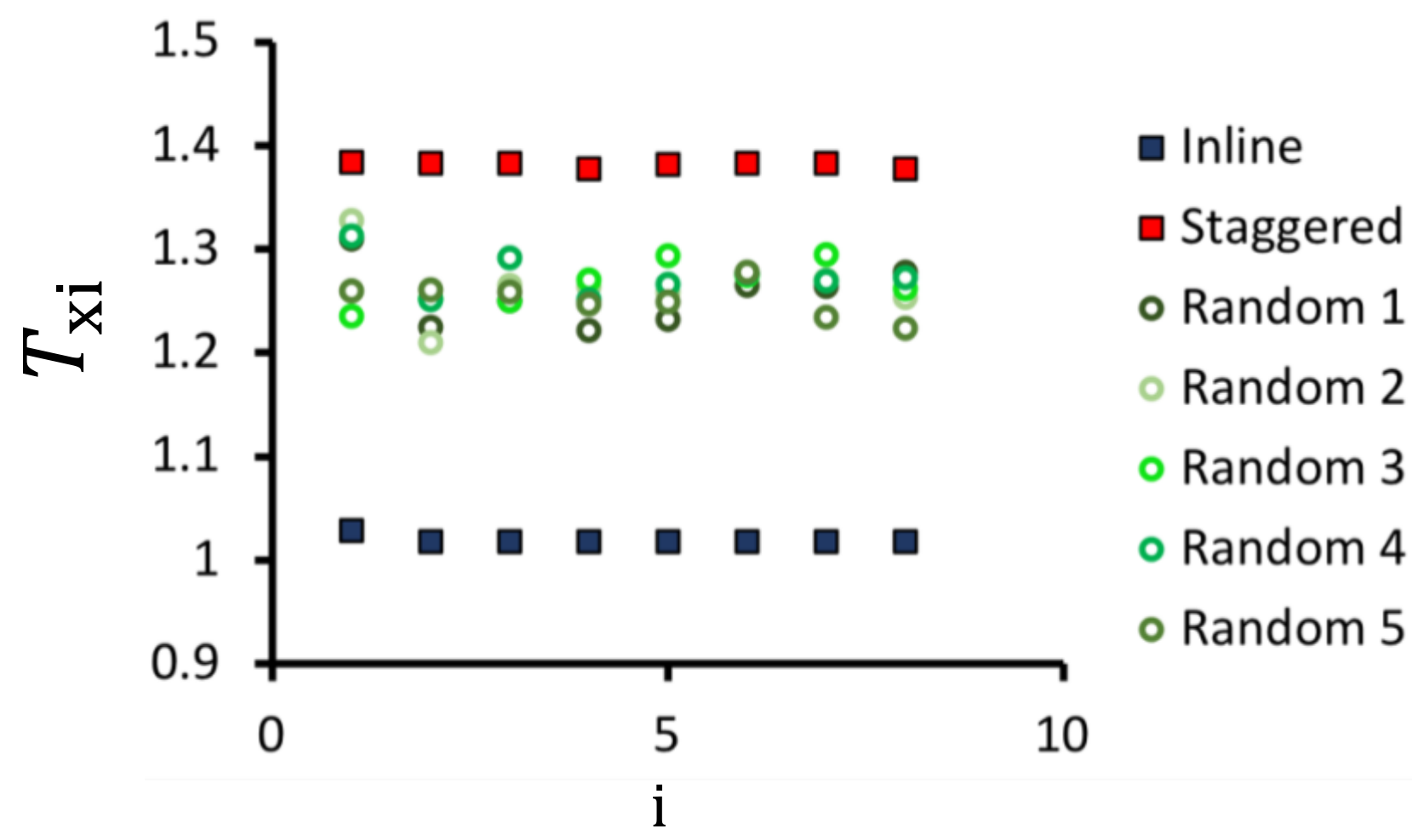

Figure 8: Quantification of structure for shocked particle laden gases; tortuosity values $T_{xi}$ for the corresponding paths numbered $i$ for seven different arrangements of particles with diameter $d_{cyl} = 0.001$ and $\phi = 5\%$.

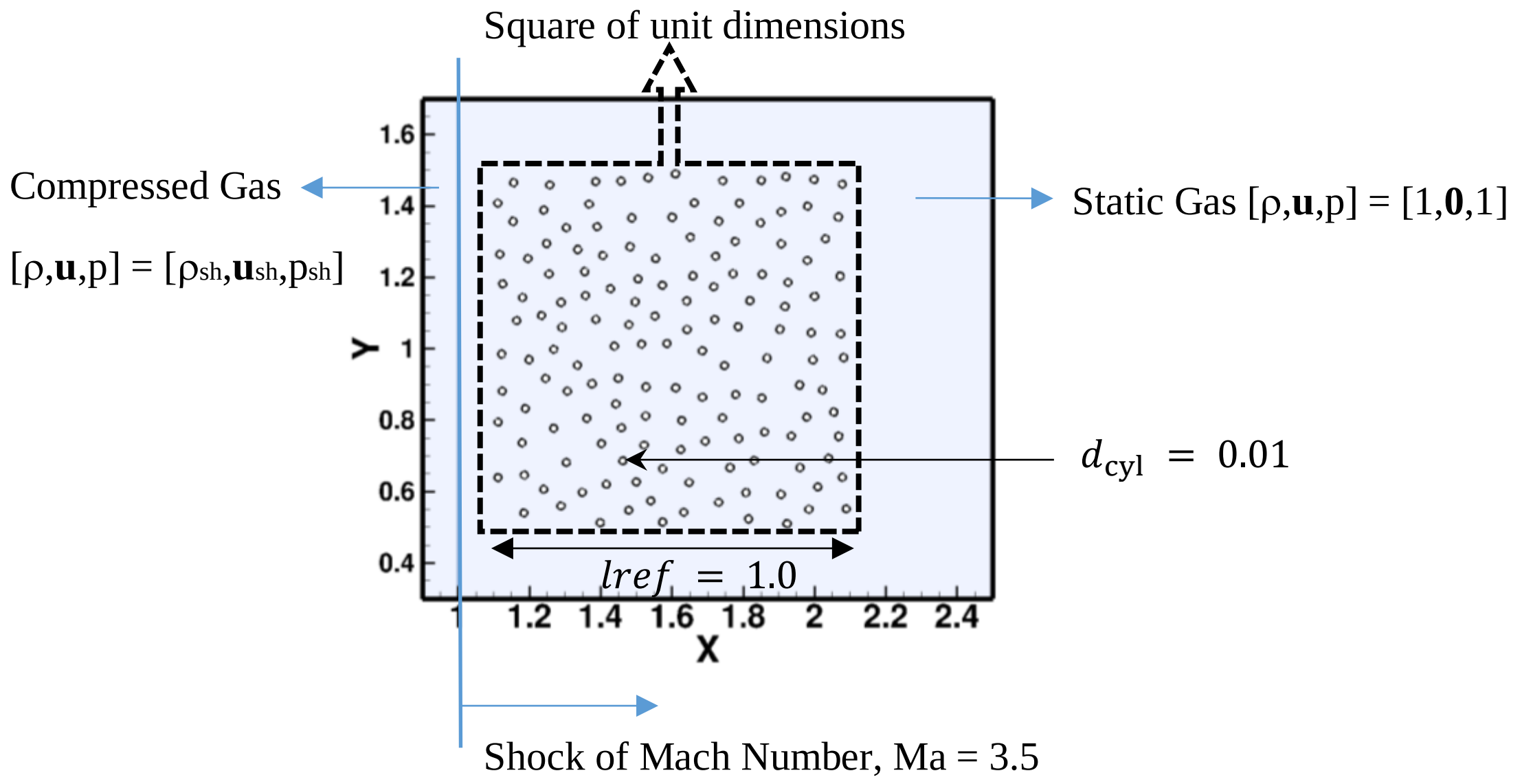

Figure 9: Schematic diagram showing the mesoscale simulations of shock interacting with a cluster of particles. The Mach number of the shock Ma = 3.5. On the right side is the static fluid with γ = 1.4, and unit density and pressure. On the left side of the shock the properties are governed by Rankine-Hugoniot jump conditions. The volume fraction of the particles in the domain is given by $\phi = n\pi d^2/4$. Where, $n$ is the total number of particles in the domain.

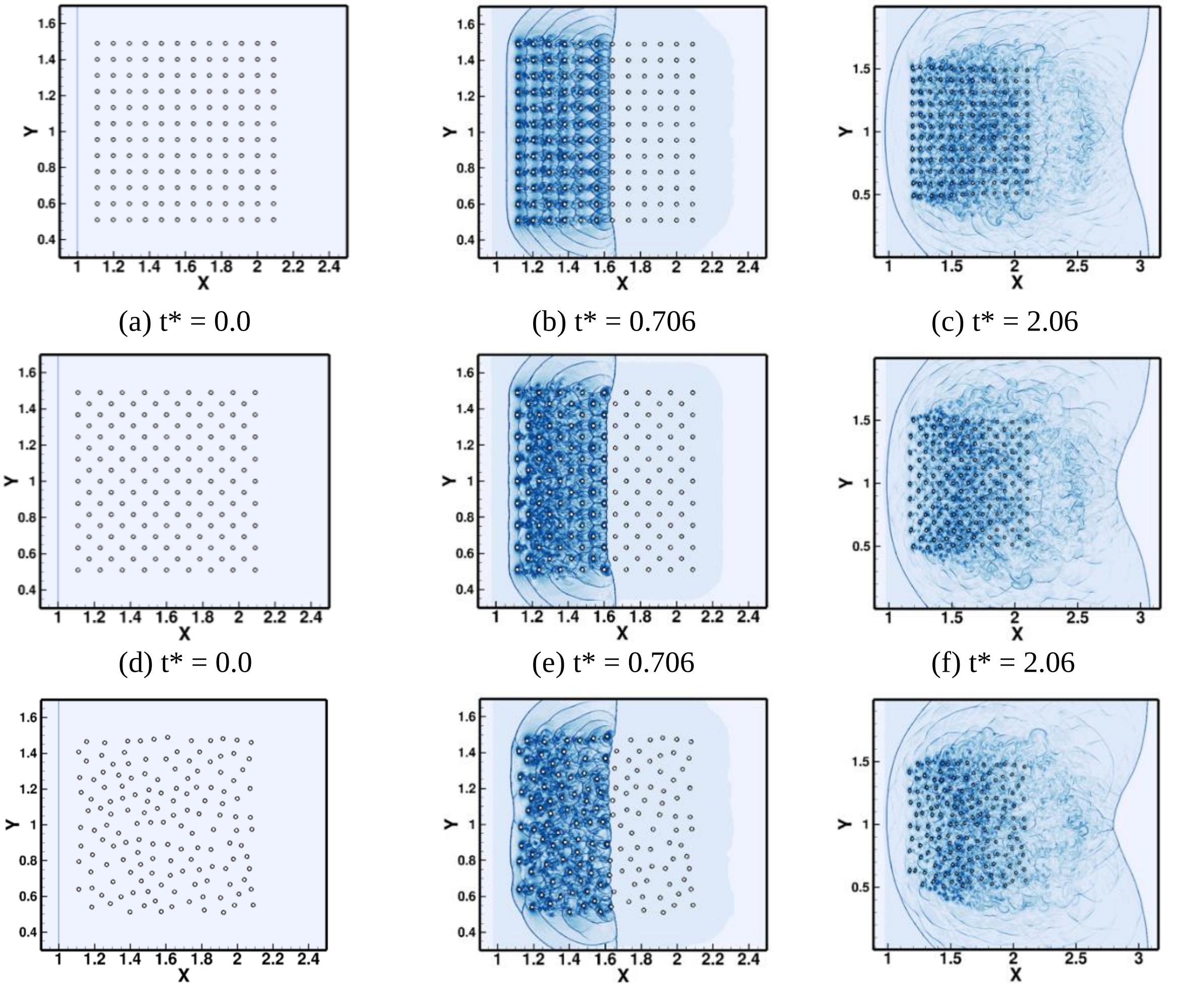

(a) t* = 0.0 (b) t* = 0.706 (c) t* = 2.06

(d) t* = 0.0 (e) t* = 0.706 (f) t* = 2.06

(g) t* = 0.0 (h) t* = 0.706 (i) t* = 2.06

Figure 10: Numerical Schlieren at various instances of time showing interaction of shock with Mach Number 3.5 with 5% volume fraction (a), (b), and (c) inline distribution of particles, (d), (e), and (f) staggered distribution of particles, (g), (h), and (i) random distribution of particles.

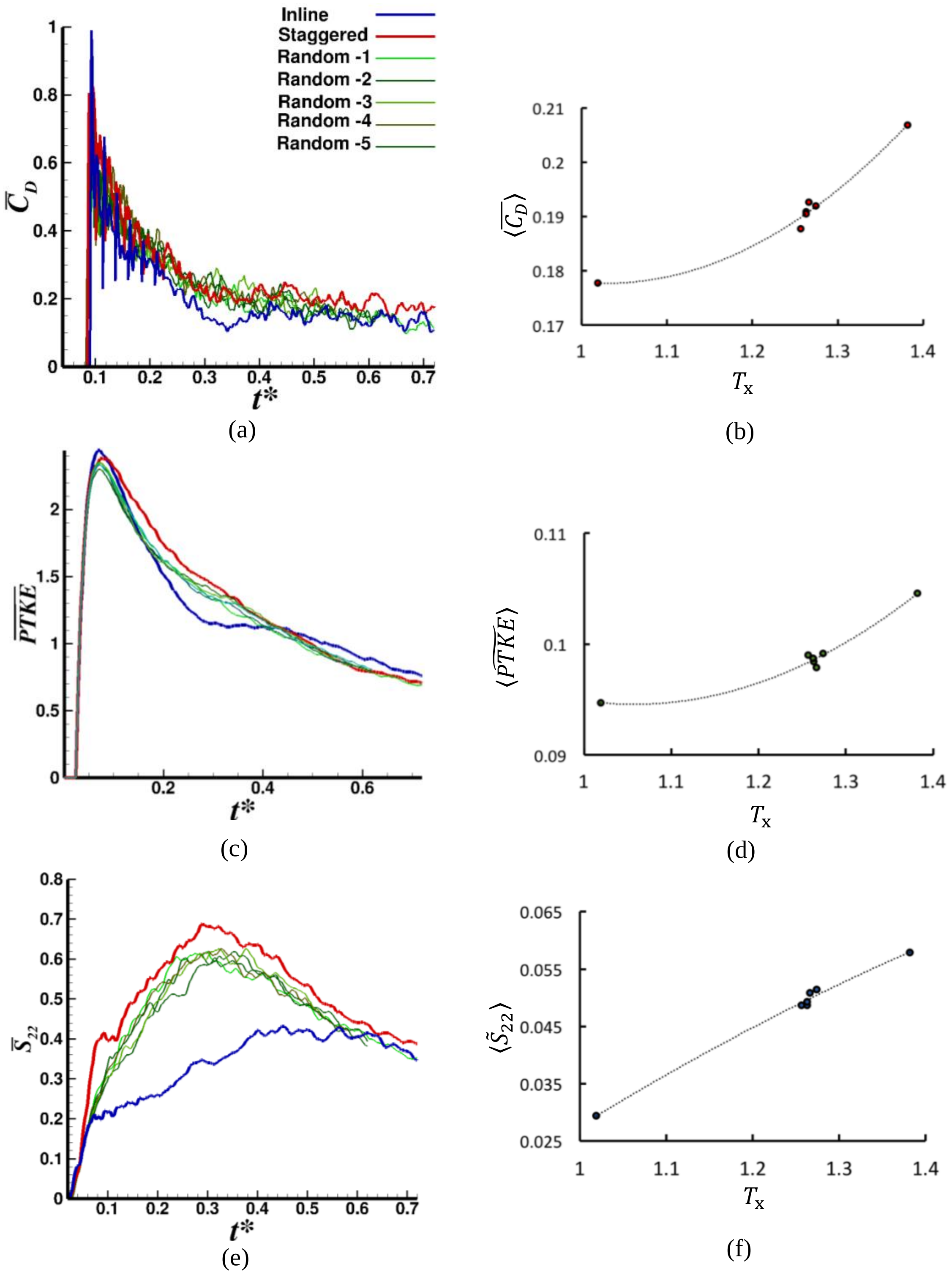

Figure 11: Evolution of mesoscale QoIs with $t^*$ in shocked gas particle flows compared for seven different arrangements (a) $\overline{C_D}$, (c) $\widetilde{PTKE}$ and, (e) $\tilde{S}_{22}$. S-P linkage established between spatial and temporal averaged mesoscale QoIs (b) $\langle\overline{C_D}\ \rangle$, (d) $\langle\widetilde{PTKE}\rangle$ and, (f) $\langle\tilde{S}_{22}\rangle$ on the y-axis and tortuosity $T_x$ on the x-axis.

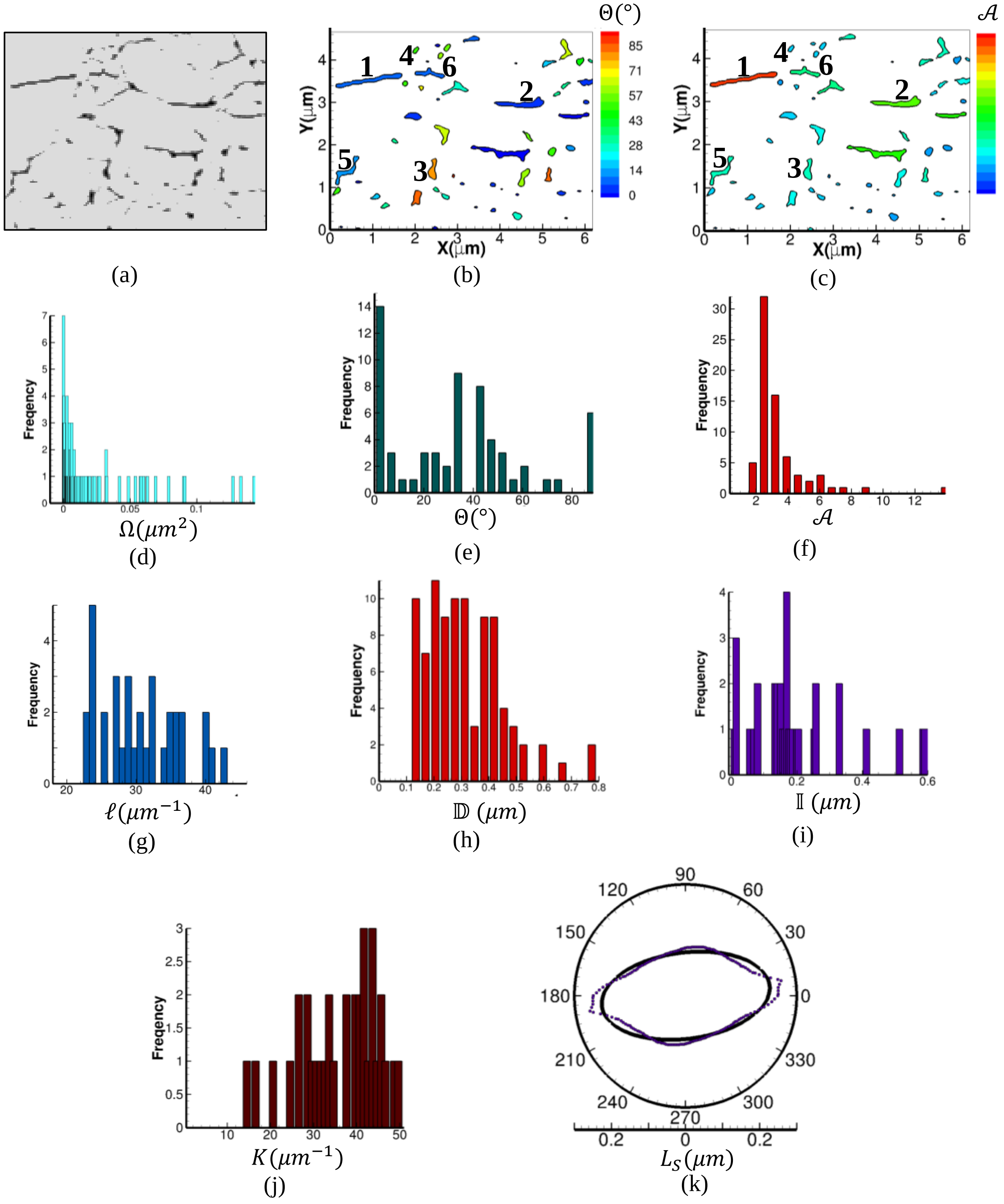

Figure 12: (a) sample A – SEM imaged microstructure of pressed HMX (courtesy Ryan Wixom, Chris Molek, and Eric Welle [1]) with dimensions 10 $\mu m \times 5 \mu m$. 2-D contour plot of (b) $\Theta$ and (c) $\mathcal{A}$ corresponding to the image in (a). Frequency distribution of structural parameters (d) void size $\Omega$, (e) void orientation $\Theta$, (f) void aspect ratio $\mathcal{A}$, (g) void interface to volume ratio $\ell$, (h) nearest neighbor $\mathbb{D}$, (i) nearest interface $\mathbb{I}$, and (j) surface curvature $K$ for the imaged microstructure corresponding to (a). (k) polar plot of the star length $L_S$ corresponding to (a).

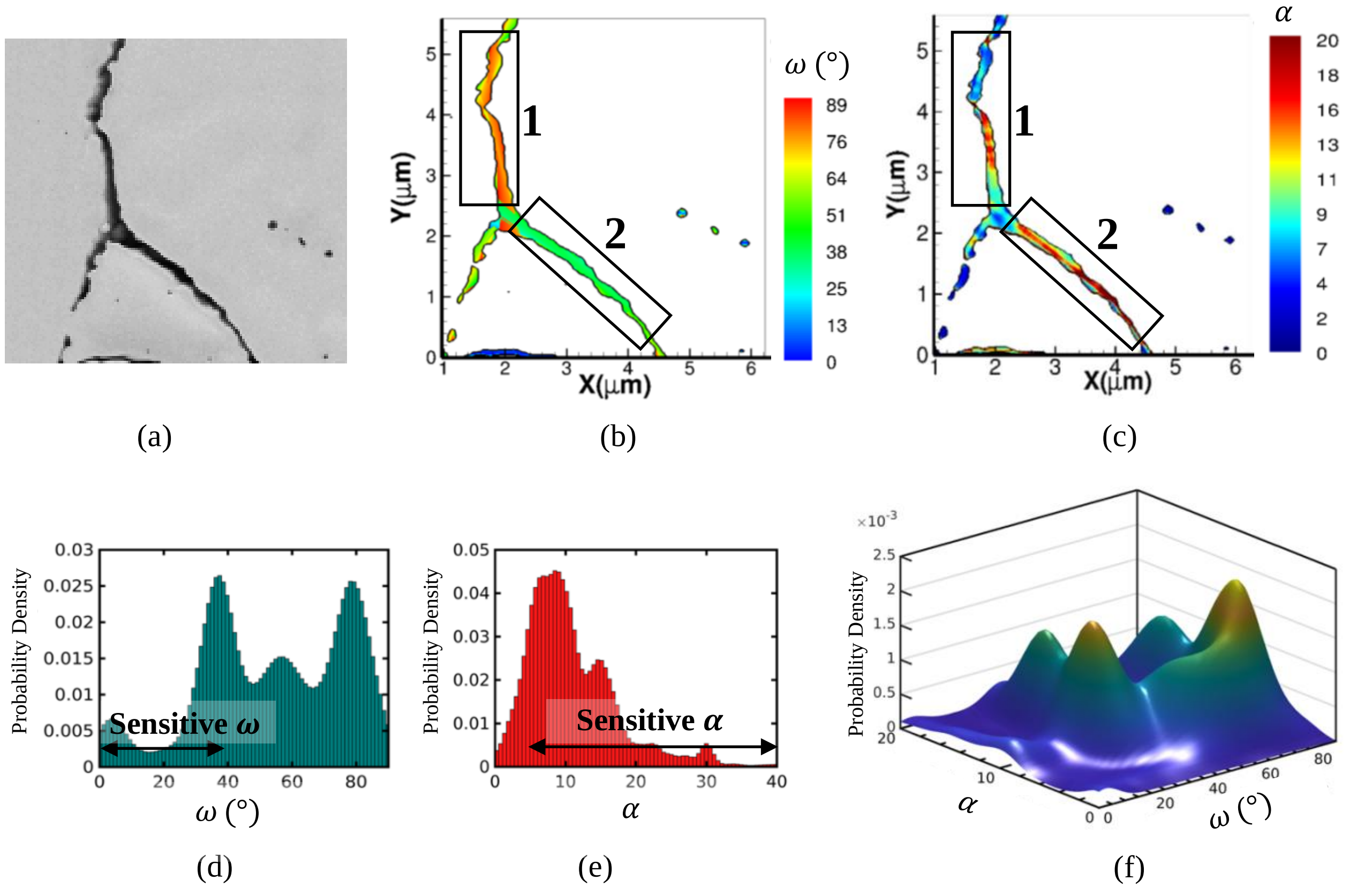

Figure 13 (a) SEM image of sample B pressed HMX [1]. Structural metrics corresponding to sample B: (b) Local orientation $\omega$ (c) local aspect ratio $\alpha$ contour, probability density distribution of (d) local orientation $\omega$ , and (e) local aspect ratio $\alpha$. (f) Joint probability distribution of local aspect ratio $\alpha$ and local orientation $\omega$ .

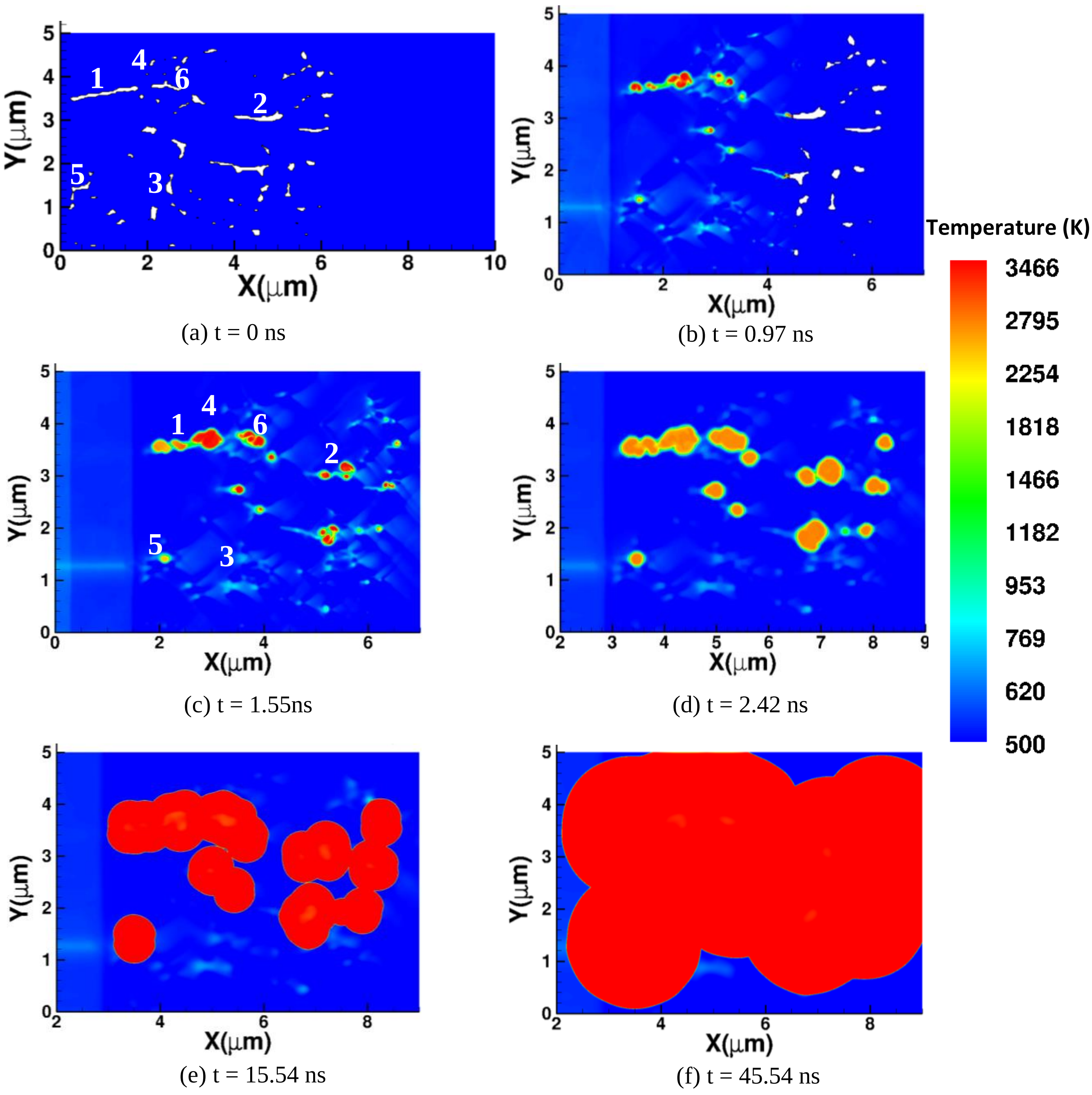

Figure 14 Temperature contour plots at various time instances obtained for shock simulation on sample A pressed HMX microstructure with a shock of particle speed = 1000 m/s and a pulse duration of 1.132 ns and a domain size of $10\mu m$ x $5\mu m$. Number of grid points used in the simulation is 2360 x 1180.

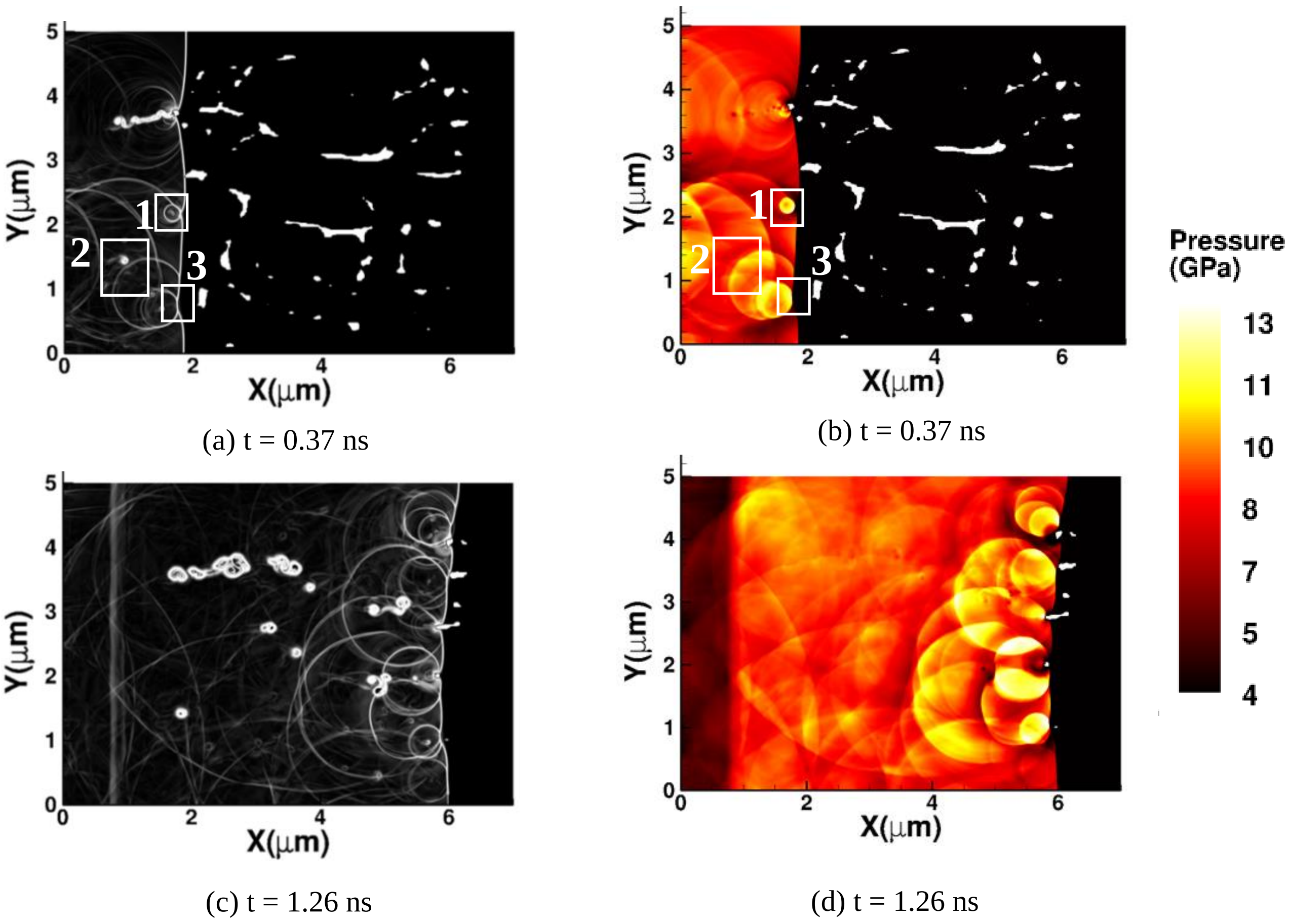

Figure 15 Schlieren contour plots at various time instances obtained for shock simulation on sample A pressed HMX microstructure with a shock strength of 650 m/s and a pulse duration of 1.132 ns and a domain size of $10\mu m$ x $5\mu m$.

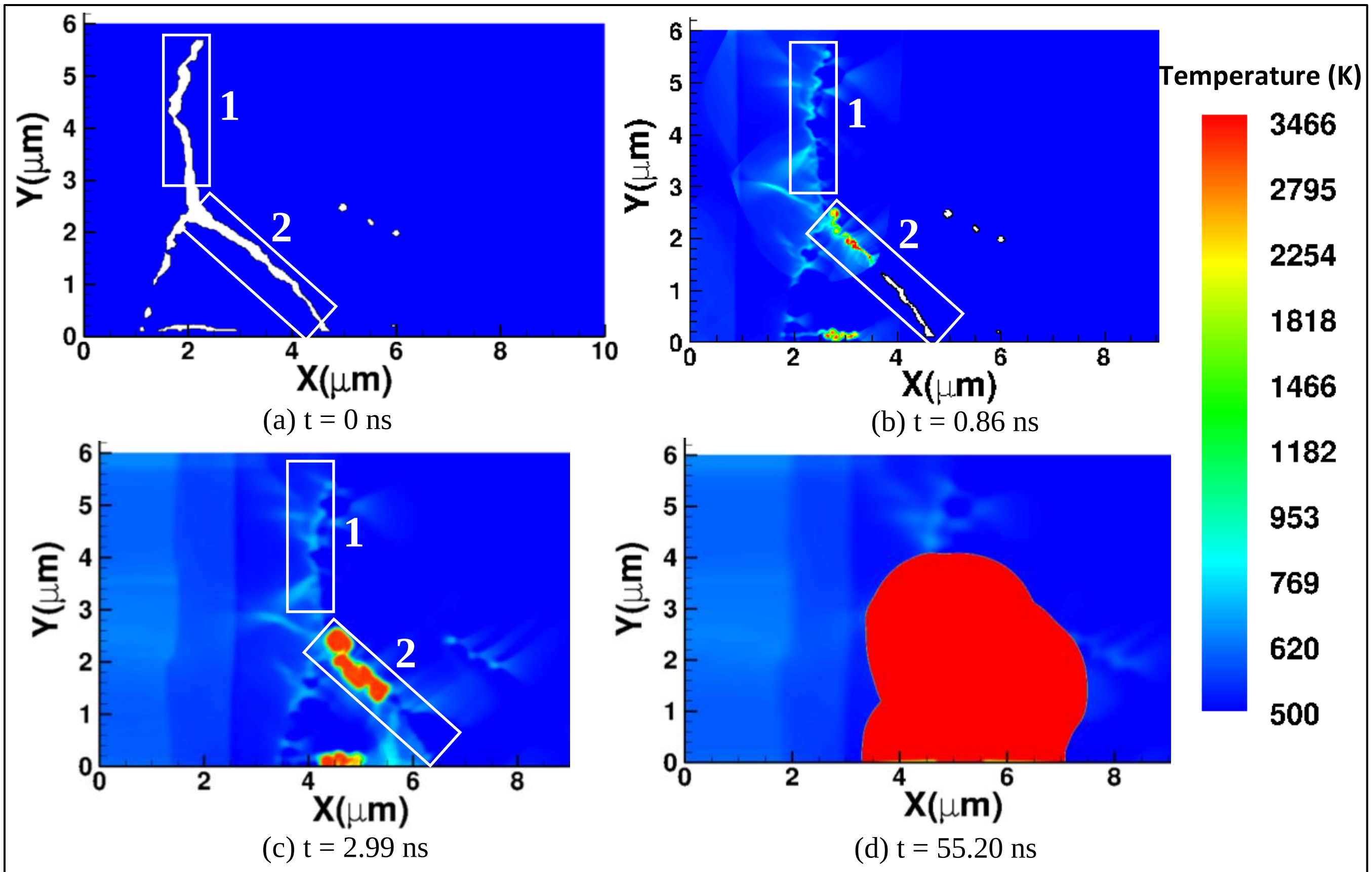

Figure 16 Temperature contour plots at various time instances obtained for shock simulation on sample B pressed HMX microstructure with a shock strength of 650 m/s and a pulse duration of 1.132 ns and a domain size of 12$\mu m$ x 6.7$\mu m$. Number of grid points used in the simulation is 2268 x 1266.

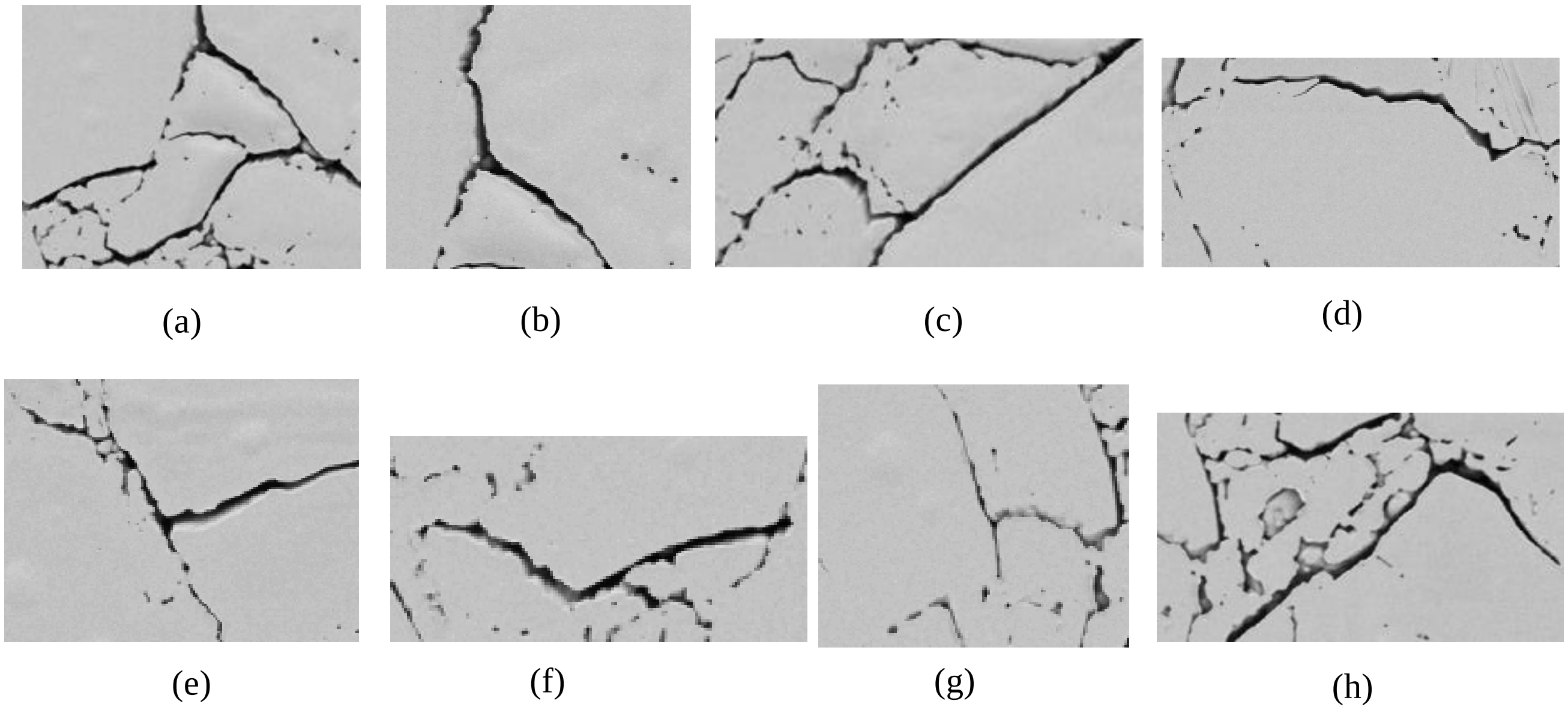

Figure 17 Class III type of SEM images of pressed HMX [1] with the corresponding dimensions. (a) Sample 1: 8.55 $\mu m \times$ 6.52 $\mu m$ (b) Sample 2: 6.22 $\mu m \times$ 5.59 $\mu m$ (c) Sample 3: 9.53 $\mu m \times$ 4.66 $\mu m$ (d) Sample 4: 9.53 $\mu m \times$ 4.70 $\mu m$ (e) Sample 5: 6.69 $\mu m \times$ 4.95 $\mu m$ (f) Sample 6: 6.31 $\mu m \times$ 3.05 $\mu m$ (g) Sample 7: 5.38 $\mu m \times$ 4.61 $\mu m$ (h) Sample 8: 10.63 $\mu m \times$ 6.27 $\mu m$.

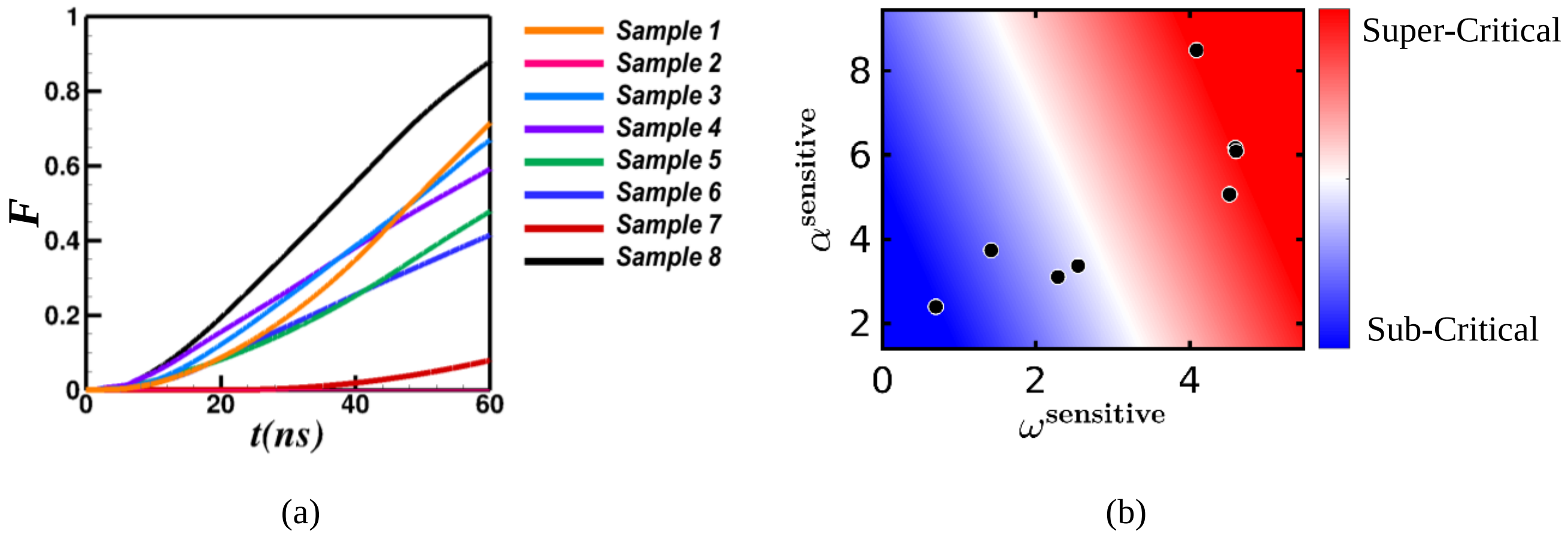

Figure 18 (a) Mesoscale QoI $F$ evolution with $t$ for the pressed HMX samples 1-8 shown in Figure 16 (a-h). (b) Super-critical and sub-critical regions predicted by the SVM based on the $\alpha^{sensitive}$ and $\theta^{sensitive}$ of the samples 1-8.